\documentclass[12pt,english]{article}

\usepackage[utf8]{inputenc}
\usepackage[T1]{fontenc}
\usepackage{lmodern}

\usepackage[top=1.5in, bottom=1.5in, left=1in, right=1in]{geometry}

\usepackage{natbib}
\usepackage{verbatim}
\usepackage{pifont}
\usepackage{amsmath,amssymb}
\usepackage{color}
\usepackage{pdfpages}
\usepackage{float}
\usepackage{graphicx}
\usepackage{setspace}
\usepackage{titling}  
\usepackage{caption}
\usepackage{multirow,booktabs}
\usepackage{array}
\usepackage{chngcntr} 
\counterwithin{table}{section}
\title{\textbf{\Large Residual Income Valuation and Stock Returns: Evidence from a Value-to-Price Investment Strategy.}\thanks{%
We gratefully acknowledge valuable comments and suggestions received from participants at various research seminars and academic conferences.}}
\author{Ahmad Haboub\thanks{%
Faculty of Business and Law, University of Northampton, Northampton, UK. E-mail address:
ahmad.haboub@northampton.ac.uk.} \\
University of Northampton \and Aris Kartsaklas\thanks{%
Department of Economics, Finance and Accounting, Brunel University London, Uxbridge
UB8 2TL, UK. E-mail address:
aris.kartsaklas@brunel.ac.uk.} \\
Brunel University London \and Vasilis Sarafidis\thanks{%
Department of Economics, Finance and Accounting, Brunel University London, Uxbridge
UB8 2TL, UK. E-mail address:
vasilis.sarafidis@brunel.ac.uk.} \\
Brunel University London}

\date{}

\begin{document}

\maketitle

\begin{abstract}
\begin{onehalfspace}

We hypothesize that portfolio sorts based on the V/P ratio generate excess returns and consist of companies that are undervalued for prolonged periods. Results, for the US market show that high V/P portfolios outperform low V/P portfolios across horizons extending from one to three years. The V/P ratio is positively correlated to future stock returns after controlling for firm characteristics, which are well known risk proxies. Findings also indicate that profitability and investment add explanatory power to the Fama and French three factor model and for stocks with V/P ratio close to 1. However, these factors cannot explain all variation in excess returns especially for years two and three and for stocks with high V/P ratio. Finally, portfolios with the highest V/P stocks select companies that are significantly mispriced relative to their equity (investment) and profitability growth persistence in the future.


        \medskip{}
		
		
	\end{onehalfspace}
\end{abstract}

\section{Introduction}\label{sec-intro}

Value-to-price (V/P) investment strategy finds its origin in the work of Frankel and Lee (1998) where the residual income valuation model is used to predict the intrinsic value. This strategy is more successful and leads to better abnormal returns over longer horizons than simple market multiples do (Ali et al., 2003; Frankel and Lee, 1998; Hwang and Lee, 2013; Goncalves and Leonard, 2023; Cong et al., 2023).\footnote{Existing empirical evidence suggests that high value-to-price stocks significantly outperform low value-to-price stocks for holding periods that extend up to three years. One explanation of this slow price convergence is the speed at which long-term fundamental information is incorporated in stock prices. An alternative explanation of the value-to-price effect is that it reflects cross-sectional risk differences.}

Over the last two decades, many researchers have tried to understand the puzzling features of value-to-price investment strategies. Frankel and Lee (1998) show that the abnormal return is not due to differences in market betas, firm size, or the book-to-market ratio. Likewise, Ali et al. (2003) conclude that value-to-price anomalies are concentrated around dates of earnings announcements.\footnote{Their findings suggest that the power to predict the returns of the V/P strategy is attributable to market mispricing and this mispricing is subsequently corrected during earnings announcement periods when a substantial amount of accounting information reaches the market. To explore the risk factors which might cause the V/P anomaly, as an alternative explanation, Ali et al. (2003) control for a large set of risk factors as suggested by Gebhardt et al. (2001) and Gode and Mohanram (2003).} Collective evidence supports the mispricing explanation of the V/P anomaly. Moreover, Hwang and Lee (2013) conclude that Fama and French's three-factor model can't explain value-to-price strategy excess returns, while the V/P factor can only explain part of it.\footnote{The intrinsic value (V) is estimated using the residual income model (Ohlson,1995; Dechow et al.,1999) and the V/P factor is constructed as a mimicking portfolio based on the V/P ratio similarly to their original factors.}
However, researchers so far have focused on first-year returns and ignored the second- and third-year portfolio returns. If the value-to-price ratio successfully predicts future returns at stock level, we hypothesize that portfolio sorts based on the V/P ratio generate significant excess returns (and consist of companies that are undervalued) for investment periods that extend up to three years.\footnote{The predictive power of the V/P strategy is comparable to that of the B/M strategy in the short term (with a one-year horizon). However, the performance of the V/P strategy significantly improved over longer horizons in comparison with those of the B/M.}
The importance of asset pricing factor models in terms of explaining value-to-price strategy returns is also explored. In particular, returns of value-to-price ratio sorted portfolios are calculated, and factor models are utilised to examine the size of excess return at both short and long investment horizons (Hou et al., 2015; Fama and French, 2015).\footnote{Fama and French (2015, 2016) used the dividend discount model to explain why profitability and investment add to the description of average returns provided by book-to-market (B/M) ratio. They found that the five-factor model largely explains the cross-sectional return patterns (related to size, B/M, profitability, and investment), the value factor becomes redundant for describing average returns and several return anomalies shrink. Hou et al. (2015) show that an empirical q-factor model consisting of the market factor, a size factor, an investment factor, and a profitability factor largely summarizes the cross section of average stock returns.}

This study contributes to the finance literature in many ways. First, fundamental stock value is calculated using the residual income model of Ohlson (1995), which combines historical information and one-year analysts' earnings forecasts (see also Feltham and Ohlson, 1995; Barth et al., 1999; Dechow et al., 1999; Myers, 1999). In addition, the valuation model here differs from those in previous studies as it allows for accounting conservatism (or book values) to affect net income, and equity prices to be estimated from a structured system of equations.\footnote{Frankel and Lee (1998) clarify that their implementation of V/P strategies is simple, and it focuses on a valuation model based on analysts' forecasts. They suggest that future research may adopt different valuation approaches that refine the model parameters. Frankel and Lee (1998) and Ali et al. (2003) used merely the financial analysts' forecasts in calculating the fundamental value, while Hwang and Lee (2013) fundament value estimates depend only on historical data.}
Second, this study provides new empirical evidence for the mispricing/risk explanation of the V/P anomaly not only at stock but also at portfolio level. In particular, the relationship between value-to-price ratio and various firm characteristics (known as risk proxies) and/or long-horizon stock returns is explored.\footnote{If the coefficient of the V/P ratio is significantly greater than zero after controlling for risk factors, it indicates that the V/P captures additional risk factors beyond the ones controlled for. In other words, it demonstrates the value-to-price anomaly at firm level.}
Finally, the ability of asset pricing factor models to explain the puzzling return patterns associated to the fundamental value-to-price (V/P) ratio is investigated.\footnote{The performance of the factor models is assessed using the Gibbons-Ross-Shanken (GRS) F-statistic. Sentana (2009) provides a survey of mean-variance efficiency tests, while Penaranda and Sentana (2015), Barillas and Shanken (2018) and Kelly et al. (2019) propose different frameworks for comparing assets pricing models and testing portfolio efficiency.}

The dataset is constructed from the merger of COMPUSTAT, CRSP, I/B/E/S for all non-financial firms listed in AMEX, NYSE, and NASDAQ during the period from 1987 to 2015. Value-to-price quintile (five) portfolios are formed. Portfolio 1 consists of firms with the lowest V/P ratio and Portfolio 5 consists of firms with the highest V/P ratio. Value-to-price portfolio returns in years one, two, and three are tested against risk factor models. Overlapping portfolio returns are also estimated, where in any given month \( t \), the V/P strategy holds a series of portfolios that are selected in the current year (\( t \)), the previous year (\( t-1 \)) and the year before that (\( t-2 \)). Finally, given that the selected stocks are held for three years, additional results are generated using dividend-adjusted monthly excess returns.

Findings reveal that high V/P portfolios outperform low V/P portfolios across horizons extending from one to three years with the performance being significantly higher at longer horizons. Overall, the V/P effect observed in this sample is very consistent with that reported in other studies (Frankel and Lee, 1998; Ali et al., 2003; Goncalves and Leonard, 2023; Cong et al., 2023). At the firm level, the relationship between long-horizon (buy-and-hold) returns and various risk proxies$-$including the V/P ratio$-$is examined. Stocks with a high V/P ratio tend to have higher idiosyncratic volatility, higher return-on-asset (ROA) volatility, and smaller size. Importantly, this study finds that the V/P ratio explains long-horizon returns even after controlling for these risk characteristics. Finally, the results show that V/P portfolio returns are associated with exposure to market, size, value, investment, and profitability risk factors. However, these factors cannot explain all variation in excess returns. Portfolios with the highest V/P stocks consist of companies that are significantly mispriced relative to their equity (investment) and profitability growth persistence in the future.

The next section of this paper is devoted to a literature review of the V/P anomaly and the mispricing versus risk explanation of it. The methodological development and data used for empirical implementation are described in Sections 3 and 4, respectively. Section 5 presents the main empirical results and discusses them. The last section presents the conclusions and offers suggestions for future research.

\section{Related Literature}\label{sec-literature}

Value strategies focus on buying stocks with low market prices relative to fundamentals such as earnings, dividends, and book values (Lakonishok et al., 1994, Asness et al., 2013; Novy-Marx, 2013). Lakonishok et al. (1994) found that value stocks yield an extra 10$\%$ return on average over glamor stocks, while this is largely due to underpricing of these stocks relative to their risk and return characteristics, and not due to being fundamentally riskier.\footnote{Investors overreact to stocks that have done very well (bad) in the past and buy (sell) them, so that these stocks become overpriced (underpriced). A value strategy buys the stocks that are underpriced and sells stocks that are overpriced, thus outperforming the market. The overreaction story is also consistent with DeBondt and Thaler (1985). However, Fama and French (1992) find that investors in value (high book-to-market) stocks earn higher average returns as a compensation to higher fundamental risk. Fama and French (1993) show that small and high book-to-market firms offer higher returns as a compensation for higher systematic risk associated with distress (more sensitive to business cycle and credit condition changes).} Novy-Marx (2013) finds that only highly profitable value stocks generate significant excess returns, while Asness et al. (2013) report consistent value return premia across eight diverse market and asset classes. Fama and French (2006) show no significant difference in the value premiums of large-cap and smaller stocks, and once they control for size and book-to-market expected returns do not seem to be related to CAPM betas. Moreover, Piotroski and So (2005) demonstrate that returns to value/glamour investment strategies are strongest among firms where expectations implied by current prices are dissimilar with the strength of their fundamentals.\footnote{Also, Daniel and Titman (1997), investigate whether portfolios with similar characteristics but different loadings on the Fama and French (1993) factors, have different returns. Once they control for firm characteristics, expected returns do not appear to be positively related to the loadings on the market, book-to-market, or size factors. Similarly, Gregory et al. (2001, 2003) highlight that there are substantial differences in returns between value and glamour portfolios that cannot be explained by their loading on the Fama-French factors.}  Lettau and Ludvigson (2001) show that value stocks earn higher average returns than growth stocks because they are more highly correlated with consumption growth in bad times, when risk premia are high.\footnote{This result lends support to the view that the reward for holding high book-to-market (B/M) stocks arises at least partly because of true non-diversifiable risk.} Petkova and Zhang (2005) provide evidence that value (growth) portfolio betas tend to covary positively (negatively) with the expected market risk premium but their covariance is far too small to explain the observed magnitude of the value premium within the conditional CAPM.

\subsection{Value-to-price strategy and stock returns}\label{subsec-stock_returns}
Frankel and Lee (1998) suggest that fundamental value-to-price trading strategy (V/P) can be used to predict cross sectional abnormal returns for up to three years. They use a version of the residual income model that incorporates financial analysts' forecasts to estimate the fundamental value (V). Their results confirm that the V/P ratio reliably predicts cross sectional stock returns, especially over longer horizons.\footnote{For 12-month horizons, the value-to-price (V/P) ratio predicts cross-sectional returns as well as the book-to-market ratio (B/M). However, over two or three-year periods, buy-and-hold returns from V/P strategies are more than twice those from B/M strategies. Thus, the V/P trading strategy is more successful and leads to higher abnormal returns than simple market multiples do.}  One explanation of this slow price convergence is the speed at which long-term fundamental information is incorporated in stock prices. An alternative explanation of the V/P effect is that it reflects cross-sectional risk differences. Frankel and Lee (1998) control for three common risk factors, namely, market beta, size, and book-to market, and find that these factors cannot explain the V/P anomaly. Therefore, the V/P anomaly may be attributed to temporary mispricing by the market, even though they do not completely rule out the possibility that V/P strategies may be riskier in other dimensions. Dechow et al. (1999) provide evidence that high value-to-price decile portfolios produce better 12-month returns compared to low ones.\footnote{Dechow et al. (1999) adopt variations of the residual income model (Ohlson,1995), which include incorporating both historical earning information and other information (or ignoring other information) and other alternatives which restrict the persistence parameters of abnormal earnings and other information either to zero or to unity in different combinations. One year ahead financial analysts' forecasts are used as a proxy for `other information' variables.}  The superior explanatory power of the simple residual income model may arise because investors overweight information in analysts' earnings forecasts and underweight information in current earnings and book value. Lee and Swaminathan (1999) and Lee et al. (1999) investigate the time-series relationship between stock price and intrinsic value and stock returns and intrinsic value, respectively. Their work emphasizes the statistical predictive ability of the V/P ratio, where value is estimated using a residual income valuation model. They claim that using a time-varying discount rate and a one-year analysts' forecast are crucial for the success of the V/P strategy.\footnote{Although this finding is consistent with market inefficiency, the authors claim they cannot rule out the possibility that the predictive power of V/P arises from time-varying expected returns. Even after accounting for well-known determinants of such risk, V/P may still capture a previously unidentified dimension of time-varying risk.}

In contrast to the previous argument, which supports the superior predictive power of the V/P ratio, Xu (2007) argues that the numerator of the V/P ratio is based on fundamental variables (book value, earnings, analysts' forecasts) which have been recognised to be correlated with future abnormal returns. Thus, it is an open question whether adding all these components to the V/P creates incremental predictive power and whether the residual income valuation model or its underlying components are the reasons for the V/P anomaly. The author concludes that the V/P has no incremental ability to explain the associated abnormal returns over its components, particularly the analysts' forecasts of earnings. Therefore, V/P has no anomalous power and the reason for the V/P effect is the investors' subjective expectations regarding its underlying variables.

Likewise, Myers (1999) and Lo and Lys (2000) raise concerns about Frankel and Lee's implementation of the residual income valuation model.\footnote{For instance, Lo and Lys (2000) argue that adding analysts' forecasts of earnings beyond one year has no significant impact on the correlation between intrinsic value and price. They claim that analysts' forecasts tend to be noisier after the first year and impounding them in residual income valuation model has insignificant effect. Instead, most of the cross-sectional correlation between price and value is primarily attributed to the book value of equity and to a lesser extent to the first year's earnings. The conclusions of Lo and Lys (2000) are in line with those of Myers (1999). An alternative argument may be that if the discount rate used to calculate the intrinsic value were too low, giving rise to high V/P, it is inevitable to obtain higher realised returns than another firm with a low V/P.}  Goncalves and Leonard (2023) find that the premium associated with the fundamental-to-market ratio (F/M) subsumes the book-to-market ratio (B/M)  premium and has been relatively stable, while the cross-sectional correlation between F/M and B/M decreased over time, inducing an apparent decline in the value premium. Cong et al. (2023) also document that value-to-price, the ratio of Residual-Income-Model based valuation to market price, subsumes the power of book-to-market ratio and generate significant returns after adjusting for common factors. Novy-Marx (2013) demonstrates that profitable firms generate significantly higher returns than unprofitable firms, despite having significantly higher valuation ratios. Controlling for profitability significantly increases the performance of value strategies, especially among the largest, most liquid stocks.\footnote{These results are difficult to reconcile with popular explanations of the value premium, as profitable firms are less prone to distress, have longer cash flow durations, and have lower levels of operating leverage.}

Ball et al., (2016,2020) reveal that cash-based operating profitability (a measure that excludes accruals) outperforms measures of profitability that include accruals (gross profitability, operating profitability, and net income). In addition, cash-based operating profitability (retained earnings-to-market) subsumes accruals (book-to-market) in predicting the cross section of average returns.

\subsection{Value-to-price strategy and mispricing}\label{subsec-mispricing}

Academics and practitioners agree that the V/P strategy can predict the cross-section of stock returns for up to three years. However, the reasons for this superior predictability of V/P strategies remain open to discussion. Frankel and Lee (1998), as noted above, turn to temporary mispricing by the market to explain the V/P anomaly, while not completely dismissing the possible riskiness of V/P strategies in other dimensions. Ali et al. (2003) investigate mispricing versus risk as the explanation of the V/P anomaly. They conclude that V/P anomalies are mostly concentrated around earnings announcement dates supporting the mispricing explanation.\footnote{Their findings suggest that the power of the V/P strategy to predict returns is attributable to market mispricing and this mispricing is subsequently corrected during earnings announcement periods, since a substantial amount of accounting information reaches the market after earnings announcement dates.}  Moreover, they observe that the V/P ratio is significantly positively associated with future abnormal returns from the V/P strategy, even after controlling for known risk factors, including book-to-market ratio, market beta, Altman's Z-score, the implied cost of capital and the debt-to-equity ratio. Unlike previous studies which focus on the general predictive ability of V/P strategy, Johnson and Xie (2004) investigate the movement of stocks in the extreme V/P quantile portfolios.\footnote{He states that if risk is the underlying reason for the V/P anomaly, then the abnormal returns of this strategy should be concentrated in the portfolio of stocks that remain in the extreme V/P portfolios. However, if mispricing is the underlying reason for the V/P anomaly, then the abnormal returns of the V/P strategy should be concentrated in the subsample of stocks in the extreme V/P portfolios that display price convergence.}  They find that less than 30$\%$ of the stocks in the extreme V/P quantiles exhibit price convergence to fundamental values after 36 months and the abnormal returns of the V/P strategy are mainly driven by this small subsample of stocks. Their empirical evidence supports the mispricing explanation of the V/P anomaly, while analysts' forecast revisions are not the driving force of price discovery for the portfolio of stocks that exhibit price convergence. Wei and Zhang (2007) argued that the V/P anomaly can be used as a good example in investigating the impact of arbitrage on the realized abnormal returns.\footnote{Shleifer and Vishny (1997) argue that when market price diverges far from fundamental value, the arbitrage becomes ineffective.}  They show that the profitability of the V/P strategy is concentrated in stocks with low arbitrage risk, whereas it is extremely weak in stocks with extremely high arbitrage risk. Their results confirm the mispricing explanation of V/P strategy.\footnote{Wei and Zhang (2007) use different measures of arbitrage risk (accrual quality, divergence of opinion, investor sophistication, firm age, idiosyncratic return volatility, liquidity, institutional ownership) and find that when stocks with any of these risks in the highest quintiles are excluded from analysis, the profitability of V/P strategies improves markedly. Their evidence confirms the view that there are limits to arbitrage in that high fundamental risk, high noise trader risk, and high transaction costs deter arbitrage activities and therefore prolong the process of stock prices to converge to their fundamental values and lower the arbitrage returns.}  Golubov and Konstantinidi (2019) study the value premium using the multiples-based market-to-book decomposition and highlight that the market-to-value component drives all of the value strategy return, while the value-to-book component exhibits no return predictability in either portfolio sorts or firm-level regressions.\footnote{Equally, Jaffe et al. (2019) demonstrate that the market-to-value component, but not the value-to-book one, predicts abnormal returns for up to 5 years and provides incremental information relative to existing asset pricing models.}

\subsection{Value-to-price strategy and risk}\label{subsec-risk}

Although existing evidence supports the mispricing explanation of V/P strategies, none of them rules out completely the possibility that the stocks in the top V/P portfolio may in some dimensions be riskier than stocks in the bottom V/P portfolio (Frankel and Lee, 1998; Ali et al., 2003). Other researchers have suggested that the mispricing explanation of V/P ratio may be premature (Lo and Lys. 2000; Myers, 1999; Kothari, 2001; Beaver, 2002).

For instance, Frankel and Lee (1988) state that the V/P anomaly could still be due to unidentified risk factors other than book-to-price ratio, firm size, and market beta. Kothari (2001) argues that the V/P strategy is quite puzzling because it generates low abnormal returns in the first year and a half, but larger abnormal returns for the next year and a half. However, inferences about long-term market mispricing over a longer period are usually confounded by omitted risk factors, the long-term nature of the anomaly itself, or other biases such as survival, statistical and performance assessment.\footnote{Beaver (2002) claims that it is very challenging to resolve the contradiction between the rapid market reaction to new information, which implies market efficiency, with the persistence of abnormal returns for three years after forming portfolios (the V/P anomaly is an example), which implies that market inefficiency is responsible.}

Hwang and Lee (2013) also investigated whether the V/P anomaly is better explained by market inefficiency or risk factors.\footnote{The Fama-French three-factor and a four-factor model, where the fourth V/P factor is constructed as a mimicking portfolio based on the V/P ratio, are used.}  They exhibit that the Fama-French three-factor model is unable to explain value-to-price strategy excess returns. Moreover, their V/P factor loading is still able to predict returns after controlling for V/P characteristics and conclude that Frankel and Lee's effect (1998) may be driven by risk factors instead of temporary mispricing. Their findings suggest that using the word ``anomaly'' to refer to V/P strategies may be inappropriate (see also Xu, 2007).

Petkova (2006) demonstrates that shocks to the aggregate dividend yield, term spread, default spread, and one-month Treasury-bill yield explain the cross section of average returns better than the Fama-French model.\footnote{When the innovations in the predictive variables are present in the model, loadings on HML and SMB lose their explanatory power for the cross-section of returns. More, the value factor proxies for a term spread surprise factor in returns, while size factor proxies for a default spread surprise factor.}  Hahn and Lee (2006) also present evidence that the size and value premiums are compensations for higher exposure to the risks related to changing credit market conditions (default spread) and interest rates (term spread), respectively.

\section{Methodology and empirical implementation}\label{sec-methodology}

First, the empirical implementation of the residual income model is discussed, while its theoretical features are briefly outlined in the appendix. This section presents the regression model used to investigate the relationship between value-to-price ratio and risk proxies (and value-to-price and long-horizon returns). Finally, asset pricing factor models are used to examine the performance of value-to-price (V/P) trading strategies.

\subsection{Empirical implementation of the residual income valuation model}\label{subsec-implementation}

This study adopts the residual income valuation model as developed by Feltham and Ohlson (1995) and Ohlson (1995) and implemented by Dechow et al. (1999), Barth et al. (1999), Barth et al. (2005) and Myers (1999).\footnote{In contrast to Feltham and Ohlson (1995), Myers (1999) does not differentiate between operating assets and financial assets because a) it is difficult, if not impossible, to separate the financial assets from the operating assets and b) residual operating income and residual income are equal since the financial assets earn only the normal income.}  The model in Equation 1 consists of three forecasting equations (a, b and c) and one valuation equation (d).\footnote{A constant is included in the abnormal income forecasting equation and in the valuation equation because abnormal income on average may be different from zero.}  To ensure no arbitrage condition, clean surplus relations and the internal consistency of the model, the valuation parameters and forecasting parameters were simultaneously estimated in a system of equations. In other words, the simultaneous estimation of the model ensured one-to-one mapping between the forecasting equations and the valuation equation (Barth et al., 2005; Pope and Wang, 2005; Myers, 1999; Tsay et al., 2008; Tsay et al., 2009; Wang, 2013). Furthermore, due to the possible correlation among the error terms $\left(\varepsilon_{1,it}, \varepsilon_{2,it}, \varepsilon_{3,it}, u_{it}, \right)$ in Equations 1a-1d, seemingly unrelated regression (SUR) is employed to estimate the system of equations (Zellner, 1962; Gallant, 1975; McElroy and Burmeister, 1988; Sarafidis and Wansbeek, 2012, 2021).\footnote{Therefore, parameter estimates from the valuation equations account for the effect of allowing regression errors from each of the forecasting equations to be correlated with those in the valuation equation. For detailed information on the estimation of a set of SUR equations with panel data see Avery (1977), Baltagi (1980), Kinal and Lahiri (1990), Biorn (2004).}

Following Barth et al. (2005) and Wang (2013), the predicted market value for each firm-year is estimated using data from the past five years for all firms in the industry, excluding any firm-specific data for the target firm being predicted.\footnote{This study used five years of data such that the estimated parameters reflect the trade-off between efficiency and stationarity. The efficiency of the estimate would improve by increasing the number of years, but the parameters are likely to become nonstationary. Also, Fama and French's industry classification are used, and the sample is divided into 12 sectors.}  Thus, the prediction was strictly considered to be out of sample prediction. In other words, the parameters and errors in forecasting and valuation equation are estimated using a jackknife procedure. For instance, to estimate the parameters for firm $i$ in industry $j$ for the year $t$, the data for all firms in industry $j$ for the period from year $t-4$ to year $t$ were included except the data for firm $i$ in year $t$. Also, the parameters were firm-year-industry specific because they incorporate data updated on a yearly basis. Parameters can vary across industries as a result of differences in economic environment and accounting practices, while the level of conservatism and cost of capital associated with abnormal earnings is also allowed to change by industry.

For the purposes of this paper, different discount rates (r) are used to calculate abnormal income $NI_{it}^a$. First, a range of discount rates from 8$\%$ to 16$\%$ is used. Second, CAPM and Fama and French's three-factor model is employed to calculate the discount rate on a five-year rolling basis (Fama and French, 1997). Empirical results did not change significantly between different methods. To maintain simplicity and be consistent with other studies (Barth et al., 2005; Tsay et al., 2008), a 12$\%$ discount rate is used.
\begin{align*}
NI_{it}^a &= \omega_{10} + \omega_{11} NI_{i,t-1}^a + \omega_{12} BV_{i,t-1} + \omega_{13} \nu_{i,t-1} + \varepsilon_{1,it}; \tag{1a} \\
BV_{it} &= \omega_{22} BV_{i,t-1} + \varepsilon_{2,it}; \tag{1b} \\
\nu_{it} &= \omega_{33} \nu_{i,t-1} + \varepsilon_{3,it}; \tag{1c} \\
MV_{it} &= \alpha_0 + BV_{it} + \alpha_1 NI_{it}^a + \alpha_2 BV_{it} + \alpha_3 \nu_{it} + u_{it}; \tag{1d} \\
\alpha_0 &= \frac{(1 + r)}{r}  \frac{\omega_{10}}{1 + r - \omega_{11}}; \\
\alpha_1 &= \frac{\omega_{11}}{1 + r - \omega_{11}}; \\
\alpha_2 &= \frac{(1 + r)\omega_{12}}{(1 + r - \omega_{11})(1 + r - \omega_{22})}; \\
\alpha_3 &= \frac{(1 + r) \omega_{13}}{(1 + r - \omega_{11})(1 + r - \omega_{33})},
\end{align*}
where $BV_{it}$ is the book value of equity; $MV_{it}$ is the market value of equity; $r$ is the cost of capital; $NI_{it}^a$ is the residual income, defined as $NI_{it} - r \times BV_{i,t-1}$; $\nu_{it}$ is other information; $\omega_{11}$ is the residual income persistence parameter ($0 < \omega_{11} < 1$); $\omega_{12}$ is the conservatism parameter\footnote{This must be positive ($\omega_{12}>0$) if residual income is driven in part by understated book value instead of monopolistic power. BV will be negatively related to future abnormal earnings if the normal return on equity book value is less than the return assumed in the empirical tests. Including the equity book value in the abnormal earnings equation allows the effects of conservatism to manifest (Feltham and Ohlson,1995; Ashton and Wang, 2015) and relaxes the assumption that the cost of capital associated with calculating abnormal earnings is a predetermined cross-sectional constant. Separate industry estimation of all equations permits the level of conservatism and, at least partially, the cost of capital associated with abnormal earnings to vary by industry.}; $\omega_{22}$ is the book value persistence or growth parameter\footnote{This must satisfy the following conditions, $1<1\omega_{22}<(1+r)$, for a going concern.}; $\omega_{33}$ is the persistence parameter of other information ($0 < \omega_{33} < 1$); $\varepsilon_{1,it}$, $\varepsilon_{2,it}$, $\varepsilon_{3,it}$, and $u_{it}$ are error terms; and the subscripts $i$ and $t$ refer to the firm and year, respectively.

Two alternative approaches to estimate the other information variable ($v$) are employed. First, the procedures of Dechow et al.\ (1999) and Ohlson (2001) are followed. In particular, the other information variable ($v$) is calculated as follows:
\begin{align*}
v_t &= \mathbb{E}_t \left[ NI_{t+1}^a \right] - \omega \times NI_t^a \tag{2a} \\
\mathbb{E}_t \left[ x_{t+1}^a \right] &= f_t^a = f_t - r \times BV_t \\
v_t &= f_t^a - \omega \times NI_t^a
\end{align*}
where $\mathbb{E}_t \left[ NI_{t+1}^a \right]$ is the conditional expectation of abnormal income for the period $t+1$ based on all information available at time $t$; $f_t$ is the consensus of analysts' forecasts of expected earnings for period $t+1$; and $\omega$ is the persistence parameter of abnormal income, estimated by ignoring the other information variable in equation (1a).

Second, the approach of Bryan and Tiras (2007) is adopted to calculate the other information variable ($v$), as expressed below.
\begin{align}
f_{i,t} = \delta_0 + \delta_1 NI_{i,t} + \delta_2 BV_{i,t} + v_{i,t}, \tag{2b}
\end{align}
where $f_{i,t}$ is the consensus of analysts' forecasts for next year's earnings by firm $i$; $NI_{i,t}$ and $BV_{i,t}$ are the net income and book value of firm $i$ in year $t$, respectively; $\delta_0$, $\delta_1$, and $\delta_2$ are regression parameters; and $v_{i,t}$ is the regression residual, which also proxies for other information in equation (2b). Bryan and Tiras (2007) regress the consensus of financial analysts' forecasts directly on the fundamental variables (BV and NI). Thus, the accuracy of the model depends solely on the accuracy of the regression residual.\footnote{The approach of Dechow et al. (1999) requires the cost of capital $(r)$ and abnormal income persistency parameter $(\omega_{11})$ to be estimated before $v$. Thus, the accuracy of $v$ depends on the accuracy of both $(r)$ and $(\omega_{11})$.} Both approaches give comparable results; hence, the main analysis here depends only on Dechow et al.\ (1999).\footnote{The average values of the parameters from equations (1a)--(1d), estimated using all data, are: $\omega_{11} = 0.732$, $\omega_{12} = 0.017$, $\omega_{13} = 0.426$, $\omega_{22} = 1.05$, $\omega_{33} = 0.56$, $\alpha_1 = 1.88$, $\alpha_2 = 0.71$, and $\alpha_3 = 2.88$. These results are comparable with those reported in previous studies (Barth et al., 2005; Dechow et al., 1999; Ohlson, 1995; Wang, 2013).}

\subsection{Value-to-price, risk proxies and stock returns}\label{subsec-risk_proxies}
The risk explanation for the superior predictability of the V/P strategy is explored by evaluating the relationship between the V/P ratio and commonly used risk proxies at firm level (equation 3). These factors are primarily motivated by previous studies (Fama and French, 1989, Gebhardt et al., 2001). Frankel and Lee (1998) investigate the extent to which firm size, book-to-market ratio and firm beta explain the predictive power of the V/P strategy. Similarly, Ali et al. (2003) and Hwang and Lee (2013) control for firm characteristics which had been suggested by Gebhardt et al. (2001) and Gode and Mohnram (2003) as risk proxies. The following equation is estimated using year and industry fixed effects\footnote{This allows intercepts to vary across industries and years but restricts slope coefficients to be the same. For panel data sets that have more firms than years, a common estimation approach is to include dummy variables for each time period (to absorb the time effect) and then cluster by firm or industry (to treat the firm/industry effect). The parametric approach only works when the dependence is correctly specified, and the firm/time effects are fixed. However, if the precise form of dependence is not known, a less parametric approach may be preferred and a solution is to cluster on two dimensions (e.g. firm, time) simultaneously, given a sufficient number of clusters exists (Petersen, 2009).}
\begin{align*}
V/P = \beta_0 + \beta_1 \text{Beta} + \beta_2 \text{Ivolatility} + \beta_3 D/M
&+ \beta_4 \ln(\text{ME}) + \beta_5 \text{Analysts} \notag \\
&+ \beta_6 \text{Altman's Z} + \beta_7 \text{Std(ROA)} + \beta_8 B/M + \varepsilon \tag{3}
\end{align*}
where $V/P$ is the value-to-price ratio; \text{Beta} is a measure of systematic risk; \text{Ivolatility} is a measure of unsystematic or idiosyncratic risk; $D/M$ is the long-term debt-to-market value ratio (leverage); $\ln(\text{ME})$ is a measure of firm size; \text{Analysts} is a measure of the financial analysts' coverage of the firm; and \text{Altman's Z} is a measure of financial distress. $Std(ROA)$ is a measure of earnings variability; $B/M$ is a measure of the book-to-market ratio (see Appendix for details on the risk proxies used here).

The relationship between the value-to-price ratio and long-horizon returns, after controlling for various risk proxies, is also examined. Long-horizon returns ($\text{Ret36}$) are the buy-and-hold returns over 36 months beginning in July of year $t$. The risk proxies, including the V/P ratio, are firm characteristics that potentially explain the cross-section of returns measured over a three-year period. In other words, these characteristics are candidates for predicting buy-and-hold returns beyond one year at the firm level. Equation (4) is estimated using year and industry fixed effects:
\begin{align*}
\text{Ret36} = \beta_0 + \beta_1 V/P + \beta_2 \text{Beta} + \beta_3 \text{Ivolatility}
&+ \beta_4 D/M + \beta_5 \ln(\text{ME}) + \beta_6 \text{Analysts} \notag \\
&+ \beta_7 \text{Altman's Z} + \beta_8 \text{Std(ROA)} + \beta_9 B/M + \varepsilon \tag{4}
\end{align*}

If the coefficient on the V/P ratio ($\beta_1$) is significantly greater than zero after controlling for various firm characteristics, it indicates that the V/P ratio captures additional risk attributes beyond those considered in the model. In other words, it signals a value-to-price anomaly.

\subsection{Value-to-price portfolio returns and factor models}\label{subsec-portfolio_returns_factor_models}
To investigate further the risk explanation of the V/P effect, this study tests V/P portfolio returns using the CAPM and Fama and French's three- and five-factor models. The purpose is to uncover whether the five-factor model explains the excess returns of value-to-price portfolios better than the three-factor model and the CAPM.

First, the CAPM is estimated by regressing monthly excess returns of the V/P quintile portfolios against excess returns of the overall market index, as shown in Equation (5a), where $R_{it}$ are monthly equally weighted returns of quintile portfolio $i$; $R_{mt}$ are monthly returns of the market index; and $R_{ft}$ are the monthly risk-free rate on Treasury bills:
\begin{align*}
(R_{it} - R_{ft}) &= \alpha_i + \beta_i (R_{mt} - R_{ft}) + \varepsilon_t \tag{5a} \\
(R_{it} - R_{ft}) &= \alpha_i + \beta_i (R_{mt} - R_{ft}) + s_i \text{SMB}_t + h_i \text{HML}_t + \varepsilon_t \tag{5b} \\
(R_{it} - R_{ft}) &= \alpha_i + \beta_i (R_{mt} - R_{ft}) + s_i \text{SMB}_t + h_i \text{HML}_t + r_i \text{RMW}_t + c_i \text{CMA}_t + \varepsilon_t. \tag{5c}
\end{align*}

Second, Fama and French's three-factor model is assessed by regressing excess returns of the V/P quintile portfolios against excess return on the market index and returns on the size (SMB) and value (HML) mimicking portfolios, as outlined in Equation (5b). The Small Minus Big (SMB) and High Minus Low (HML) book-to-market mimicking portfolios are formed by independently sorting all stocks in NYSE, AMEX, and NASDAQ into two stock size portfolios (Small and Big) and three book-to-market portfolios (Low, Medium, and High).

Third, a five-factor model is estimated where excess returns of value-to-price portfolios are regressed against excess returns on the market index and returns on size (SMB), value (HML), profitability (RMW), and investment (CMA) mimicking portfolios, as shown in Equation (5c). Returns on Robust Minus Weak (RMW) operating income and Conservative Minus Aggressive (CMA) investment mimicking portfolios are calculated in a similar way to the returns on the HML portfolio.

Finally, the performance of the factor models is compared using the F-statistic of Gibbons et al.\ (1989), also known as the GRS F-statistic.\footnote{Sentana (2009) provide a survey of mean-variance efficiency tests, while Penaranda and Sentana (2015) propose a unifying framework for the empirical evaluation of asset pricing models. Moreover, Barillas and Shanken (2018) and Kelly et al. (2019) use a Bayesian framework and Instrumental PCA methods, respectively, to compare asset pricing models.} The null hypothesis of this test proposes that the intercepts $\alpha_i$ are jointly equal to zero. In other words, if the intercept in the regression of value-to-price portfolios' excess returns against the asset pricing factors does not differ significantly from zero, then the asset-pricing model adequately captures the expected returns of value-to-price portfolios. Otherwise, it indicates a V/P anomaly at the portfolio level.

\section{Data\label{Data}}

The dataset used in this study consists of all AMEX, NYSE, and NASDAQ non-financial firms from the merger of the COMPUSTAT fundamental files, CRSP returns files, and Thomson I/B/E/S summary files of analysts' forecasts for one year ahead.

For a firm to be included in the equity valuation estimate, it must satisfy the following conditions. First, it must have valid data for its book value, net income before extraordinary items, outstanding shares, and fiscal year closing price from the fundamental COMPUSTAT files; and one-year-ahead consensus forecasts by financial analysts for earnings per share (EPS) from the Thomson I/B/E/S summary files.

Second, the firm must have total assets of at least $\$10$ million and a closing share price greater than one dollar, to mitigate the effect of small companies and ensure a stable V/P ratio.\footnote{Frankel and Lee (1998) claim that firms with stock price of less than one dollar are characterized by an unstable V/P ratio and poor market liquidity.}

Third, firms with negative book value and/or negative consensus in the financial analysts' forecasts for one year ahead were deleted from the sample, because including them implied a negative market value (Bryan and Tiras, 2007).

Finally, the dataset is restricted to firms with a fiscal year ending in December, to simplify the analysis and to ensure that there is a six-month gap between the fiscal year-end and the portfolio formation date.

After applying all filters, the final sample used to estimate fundamental values consists of 22,873 firm-year observations over the period 1987--2015. Table 1A (Appendix) displays the distribution of firms in the sample by industry and year. Interestingly, the number of observations in the durable goods sector was the lowest, while business equipment had the highest.

The fundamental value for each firm-year observation is estimated using the previous five years of accounting data. At the end of June each year, all stocks are sorted into five portfolios based on the value-to-price ratio. Portfolio 1 consisted of stocks with the lowest V/P ratio, while stocks with the highest V/P ratio were in Portfolio 5. The fundamental value of December in year $t-1$ is matched to the share price for June in year t to calculate the value-to-price ratio and form the corresponding portfolios. This procedure to ensures that the accounting variables were known before the returns were calculated. For a firm to be included in the V/P portfolios, the monthly return data had to be available from July in year $t$ to June in year $t+1$. Monthly returns were collected from the CRSP monthly files for the whole sample period. After matching the estimated fundamental value with the monthly return data, firm-year observations were reduced to 16580 over the period between 1993 and 2015. For comparability reasons, two additional trading strategies based on book-to-market and equity market value (size) are considered. Like the V/P trading strategies, at the end of June each year all stocks were sorted by B/M or ME into five portfolios. For a stock to be included in the portfolio, the return data had to be available, at least for the next 12 months from the portfolio formation date. Equal weighted returns and size adjusted ones are calculated across horizons of one, two and three years by compounding the monthly return data for each of the quintile portfolios. These and other characteristics, for the sake of comparability are reported below in Table 2.

\section{Results and Discussion}\label{sec-results}

\subsection{Portfolio returns and characteristics}\label{subsec-portfolio_returns_characteristics}

Table 1 reports the characteristics of the quintile portfolios formed by the market equity (ME), book-to-market ratio (B/M), and value-to-price ratio (V/P). All firms in the sample are divided into five quintile portfolios at the end of June each year based on one of these measures at a time. Table 1 provides the average ME, B/M and V/P value for each portfolio, as well as the average post-formation market beta and the average raw/size-adjusted buy and hold returns over the next 12 months (Ret12/SRet12), 24 months (Ret24/Sret24) and 36 months (Ret36/SRet36).\footnote{The post-market beta for each firm is calculated by regressing the market index against the contemporaneous firm-monthly returns over the next 36 months. The size-adjusted buy and hold returns were calculated as the difference between the raw buy and hold returns and the corresponding CRSP size-decile index returns.}  The purpose of calculating the size-adjusted buy and hold returns is to control for the effect of size differences (ME) among the quintile portfolios. The number of observations for each portfolio is reported in the last row of each panel of Table 1 and applies to all variables except the post estimation returns. The last column of Table 1 reports the post formation returns for the hedge portfolios. Hedge portfolios are formed by taking a long position in portfolio Q5 and a short position in portfolio Q1. Statistical significance of the difference (Q5-Q1) is assessed by computing portfolio characteristics on a yearly basis. Finally, time-series variations of the estimated value are used to compute the statistical significance for the mean value over the whole sample period.\footnote{The procedure proposed by Newey and West (1987) was used to correct for the serial correlation in buy and hold returns which was induced by overlapping the holding periods beyond the first year (Ret24/SRet24 and Ret36/Sret36).}

Panel A of Table1 displays that a hedge portfolio, formed by taking a long position in large ME stocks and a short position in small ME stocks, generates an average row (size adjusted) buy-and-hold returns of $-7.8\%$ ($-6.3\%$), $-16.8\%$ ($-10.6\%$) and $-28.6\%$ ($-15.1\%$) over 12-, 24- and 36-months period respectively. These results indicate that firms with smallest ME mostly outperform firms with largest ME. The portfolio of small stocks seems to have higher beta risk compared to large stocks (1.262 vs. 1.003), while their book-to-market ratios are higher than one indicating that small stocks are price low relative to their book values. Similarly, the value-to-price ratio for small stocks is higher than two demonstrating that the book value plus the discounted future abnormal earnings is more than twice their market values.

Results for the book-to-market sorted portfolios are reported in panel B of Table 1. The firms in Q1 (the lowest B/M ratio) earn on average raw (size-adjusted) buy and hold returns of $13.6\%$ ($2.4\%$) over a one-year horizon, while the firms in Q5 earn $17.7\%$ ($5.4\%$). The difference of $4.1\%$ ($3\%$) is statistically significant at $5\%$ and is comparable in magnitude to the findings in Frankel and Lee (1998). Results also suggest that the B/M effect is true over longer horizons. For instance, the B/M hedge portfolio (Q5-Q1) generates on average raw (size-adjusted) buy and hold returns of $15.6\%$ ($10.6\%$) over the next 36-month period, which is statistically significant at $5\%$. These results confirm the B/M effect widely documented in the finance literature (Lakonishok et al., 1994). Moreover, exposure to market risk (beta) does not explain the difference in returns between value stocks and glamour stocks. In addition, the value-to-price ratio is slightly lower than 1 for glamor stocks but significantly higher than one for value stocks.

Returns and characteristics of value-to-price sorted portfolios are reported in panel C of Table 1. Portfolios formed by value-to-price ratio look like those formed by book-to-market ratio. First, firms in the lowest V/P quintile (Q1) have the lowest B/M ratio, while firms in the highest V/P quintile (Q5) have the highest B/M ratio. In other words, the B/M and V/P ratios are positively correlated with each other. More importantly, a hedge portfolio formed by the V/P ratio produced on average raw (size adjusted) buy and hold returns of $5.3\%$ ($3.2\%$), $13.7\%$ ($8.8\%$) and $27.8\%$ ($14.5\%$) over the next 12 months, 24 months, and 36 months, respectively. Results indicate that the predictive power of the V/P strategy is comparable to that of the B/M strategy in the short term (with a one-year horizon). However, the performance of the V/P strategy significantly improved over longer horizons in comparison with those of the B/M. For instance, the performance of the V/P hedge portfolio spread over 36 months was $27.8\%$ ($14.5\%$), compared with only $15.6\%$ ($10.6\%$) for the B/M hedge portfolios. Likewise, market risk does not explain the difference in returns between high and low value-to-price portfolio returns. Interestingly, portfolios with the highest value-to-price ratio also have the lowest market value of equity but the size effect explains only part of the difference in returns and primarily in the first year. Overall, the V/P effect reported in this study is very consistent with that reported in other studies (Frankel and Lee,1998; Ali et al., 2003, Cong et al., 2023).

\begin{center}
\textbf{[Insert Table 1]}
\end{center}

\subsection{Value-to-price, risk proxies and stock returns}\label{subsec-risk_proxies_stock_returns}
As noted above, to investigate the relationship between the V/P ratio and risk as well as the V/P and long-horizon returns, various traditional firm risk proxies were used (Frankel and Lee, 1998; Ali et al., 2003). Particularly, the relationship between value-to-price ratio and firm characteristics is examined according to equation 3.\footnote{The B/M ratio is included as a proxy for firm growth, while Beta and Ivolatility capture the systematic and non-systematic risks of stock variability. Size and Analysts were used to capture the differences in the information environment and their impact on the risks perceived among small and large firms. In addition, Altman's Z-score is included to capture the risk of financial distress, the D/M ratio captures the influence of firm leverage, and the standard deviation of ROA (Std. ROA) is a proxy of firms' earning variability. }  The Pearson (Spearman) correlation matrix among the V/P and various risk proxies is reported in Table 2. Results exhibit that the V/P ratio is positively and significantly correlated with B/M, beta, Ivolatility, D/M and Std. ROA, which indicates that the firms with high value-to-price ratio were also the ones with high firm-specific risk characteristics. However, the negative and strong association of value-to-price with size and analyst coverage indicated that the mispricing (or risk) is higher among small firms and firms with low analyst coverage.

\begin{center}
\textbf{[Insert Table 2]}
\end{center}

To investigate the risk explanation of the V/P effect, Table 3 reports the regression analysis results for Equation 3, while two different versions of the model are estimated. In the first one, the V/P ratio is regressed against Beta, Ln (ME) and the B/M ratio, as suggested by Frankel and Lee (1998). It is clear, from the first two columns of Table 3, that the coefficients on Beta and B/M are positive (0.144 and 0.027, respectively) and statistically significant (t-statistics of 8.72 and 6.51, respectively), while the coefficient on Ln(ME) is negative and statistically significant (t-statistic of -43.03). Results exhibit that firms with high value-to-price ratio are characterised by a high book-to-market ratio, high beta, and small size, while firms with low value-to-price ratio are large firms with low beta and low book-to-market ratio. Findings are also in line with those of Frankel and Lee (1998). In the second model, which includes all risk factors, results are reported in the last two columns of Table 3. The coefficient on B/M, Ivolatility, Std. (ROA) and D/M are positive and strongly significant, which indicates that firms with a higher V/P ratio are riskier and likely to require a higher expected return. The negative and significant coefficient on Ln(ME) suggests that a higher V/P ratio is associated with smaller firms, which support the risk explanation of the V/P strategy. The sign of the coefficient on Beta, Analysts and Altman's Z-score are not consistent with the risk explanation of the V/P strategy. Notably, the coefficient on Beta is positive and significant in the first model but become negative and significant in the full model. The positive sign on Analysts and the negative sign on Altman's Z-score indicate that firms with a high V/P ratio are less likely to be firms with substantial risk of bankruptcy (Z-score) or high liquidity risk (low analyst coverage). Overall, results of the first model support the risk explanation of the V/P effect, with firm size and beta being highly corelated with the value-to-price ratio. Furthermore, the regression results of the full model indicate that stocks with a high V/P ratio also have higher idiosyncratic volatility, higher return-on-asset (ROA) volatility and smaller size. Interestingly, in the second regression it is high idiosyncratic risk rather than beta that relates to high V/P ratios. Lastly, it is not certain that the V/P effect is driven by omitted risk factors.

\begin{center}
\textbf{[Insert Table 3]}
\end{center}

In addition, the relationship between value-to-price ratio and long-horizon stock returns is explored after controlling for various risk factors, as expressed in equation 4. If the coefficient of the V/P ratio ($\beta_1$) is significantly greater than zero, conditioning on various risk proxies, it indicates that value-to-price ratio captures additional risk factors beyond the ones controlled for. Put differently, it supports the V/P anomaly. Table 4 reports the regression results for three different variations of the model in equation 4. The first two columns of Table 4 present the results of regressing three-year buy-and-hold returns (Ret36) on the value-to-price ratio as the sole explanatory variable. The positive and significant coefficient (t-statistics 10.23) confirms the V/P effect in the dataset. The second model, reported in columns 3-4, has Beta, Ln(ME) and the B/M ratio in addition to the V/P as explanatory variables. Results demonstrate that size and the V/P ratio are strongly associated with high long horizon (buy-and-hold) returns. The third model, reported in the last two columns, includes all variables of Equation 4. Results indicate that the coefficient on V/P remains significant and positive (t-statistics of 1.9 for the full model) after controlling for risk proxies; thus, confirming that omission of risk factors is not a likely explanation of the V/P effect. Also, it is important to mention that risk proxies such as idiosyncratic volatility, size, leverage, and Z-scores are significantly linked with high long-term (buy-and-hold) stock returns.

\begin{center}
\textbf{[Insert Table 4]}
\end{center}

\subsection{Value-to-price portfolio returns and risk factors}\label{subsec-portfolio_returns_risk_factors}

\subsubsection{First year and overlapping returns}
In this section, the relative performance of the Capital Asset Pricing Model (CAPM), the Fama-French three- and five-factor models is assessed in terms of explaining monthly V/P portfolio excess returns (GRS statistics for each model are also reported).\footnote{Petkova (2006) reveals that shocks to the aggregate dividend yield, term spread, default spread, and one-month Treasury-bill yield explain the cross section of average returns better than the Fama-French model. Hahn and Lee (2006) also demonstrates that the size and value premiums are compensations for higher exposure to the risks related to changing credit market conditions, default spread, and interest rates, term spread, respectively.}  Aharoni et al. (2013) uncover a positive relation between expected profitability and returns, and a negative one between expected investment and returns. Expected profitability and expected investment have a statistically significant positive and negative relation with expected returns but they don't provide an economically significant enhancement of the Fama-French model.\footnote{The authors conclude that any benefit from adding expected profitability and expected investment to size, and B/M is limited to picking the bottom performing four percent of firms.}  Golubov and Konstantinidi (2019) study the value premium using the multiples-based market-to-book decomposition and reveal that the market-to-value component drives all of the value strategy return, while the value-to-book component exhibits no return predictability in either portfolio sorts or firm-level regressions. Similarly, Jaffe et al. (2019) highlight that the mispricing component (market-to-value), but not the growth options component (value-to-book), predicts abnormal returns for up to 5 years and provides incremental information relative to existing asset pricing models.

Goncalves and Leonard (2023) find that the premium associated with the fundamental-to-market ratio (F/M) subsumes the book-to-market ratio (B/M)  premium and has been relatively stable, while the cross-sectional correlation between F/M and B/M decreased over time, inducing an apparent decline in the value premium. Cong et al. (2023) also document that value-to-price, the ratio of Residual Income Model (RIM) based valuation to market price, subsumes the power of book-to-market ratio and generate significant returns after adjusting for common factors. Additionally, Novy-Marx (2013) demonstrates that profitable firms generate significantly higher returns than unprofitable firms, despite having significantly higher valuation ratios. Controlling for profitability significantly increases the performance of value strategies, especially among the largest, most liquid stocks.\footnote{These results are difficult to reconcile with popular explanations of the value premium, as profitable firms are less prone to distress, have longer cash flow durations, and have lower levels of operating leverage.}  Ball et al., (2016, 2020) provide evidence that cash-based operating profitability (a measure that excludes accruals) outperforms measures of profitability that include accruals (gross profitability, operating profitability, and net income). Additionally, cash-based operating profitability (retained earnings-to-market) subsumes accruals (book-to-market) in predicting the cross section of average returns.

The value-to-price ratio in this study combines the stock's book value with profitability (abnormal earnings) and is consistent with other empirical studies that either use fundamental value-to-price (Goncalves and Leonard, 2023; Cong et al., 2023) strategy or a double-sort strategy based on book-to-market ratio and gross profitability (Novy-Marx, 2013). Every year at the end of June stocks are ranked in terms of value-to-price ratio. Five portfolios (from lowest to highest V/P) are formed, and their monthly returns are recorded until next June, where portfolios are rebalanced subject to the new value-to-price stock ranking (first year returns). However, existing empirical evidence and results discussed in section 5.1 reveal that high value-to-price stocks significantly outperform low value-to-price stocks for holding periods that extend up to three years. For this reason, overlapping returns are also calculated, where in any given month t, the V/P strategy holds a series of portfolios that are selected in the current year as well as in the previous year (t-1) and the year before that (t-2).

Specifically, this strategy selects stocks on the basis of current (year t) value-to-price ratio, holds them for 36 months, and at the same time closes out the position initiated in year t-3. Thus, under this trading strategy weights are revised on 1/3 of the securities in the entire portfolio in any given year and carry over the rest from the previous two years. Finally, given that the selected stocks are held for three years, additional results are produced using dividend adjusted monthly excess returns.

Panels A of Table 5 report the intercepts and slopes for five V/P quintile portfolios produced by the CAPM. First-year results show that the coefficients on the market risk premium are positive and significant for all V/P portfolios, while the intercept is not significantly different from zero for four out of five V/P portfolios. The value of the intercept is positive and weakly significant (at $10\%$) for the highest V/P portfolio only. In the case of overlapping returns, the V/P strategy yields a positive and strongly significant alpha (0.006) for the highest V/P stocks compared to the weakly significant alpha (0.002) for the lowest V/P ones. Regardless of using first year or overlapping returns, the V/P portfolios load higher on the market risk premium shifting from low to high V/P stocks, although the relation is not linear (U-shape).
Fama-French's three-factor model results are reported in panel B of Table 5. The coefficients on the market, size and B/M factors are positive and significant across the V/P portfolios except the HML of the lowest V/P portfolio (negative and significant). The loadings on the size factor increase monotonically from low to high V/P stocks, while the ones on market index display a similar U-shape to the CAPM. The highest of the V/P stocks only have the third highest loading on the value factor. The significant coefficients confirm that value-to-price excess returns vary due to the differences in size, book-to-market ratio, and market betas across quintile portfolios. Regarding the intercepts of the three-factor model, they are positive but insignificant across all V/P portfolios.

\begin{center}
\textbf{[Insert Table 5]}
\end{center}

For the overlapping returns, monthly excess returns are explained by exposure to market, size and value factors and produce insignificant alphas. Specifically, the highest of the V/P stocks always load greater on the market index and size compared to the lowest V/P ones, while their loading on book-to-market is not following this trend (third highest). However, a positive and significant alpha (0.004) is observed for the highest V/P portfolio which indicates that the three-factor model cannot explain all variation in excess returns among the highest V/P stocks. The highly significant alpha reported for the highest V/P stocks also translates into a risk-adjusted annual return of $5.6\%$ and signals the importance of holding the highest V/P stocks beyond year one. Below, two additional risk proxies are examined, namely profitability and investment, other than the market, size, and value (Hou et al.,2015; Fama and French,2015).

Panel C of Table 5 report the intercepts and slopes for the five-factor models. This model adds profitability and investment factors to the Fama-French three-factor model. Novy-Marx (2013) identifies a proxy for expected profitability that is strongly related to average return. Aharoni et al. (2013) document a weaker but statistically reliable relation between investment and average return. For first year returns, the coefficients on the five risk factors are positive and significant in most of the cases. Specifically, the market, size and B/M factors are positively related to V/P portfolio excess returns with loadings showing an increasing, but not monotonic pattern. The coefficients on the profitability (RMW) and investment (CMA) factors are positive and significant only for the V/P portfolios in the middle (Q2, Q3, Q4). Interestingly, the profitability and investment factors do not seem to explain excess returns of the highest and lowest V/P portfolios. In Novy-Marx (2013) stocks with high B/M and high gross profit also realize high stock returns. Given that the V/P ratio is equivalent to adding the present value of future abnormal profits on the nominator of the B/M ratio, the portfolio with the highest V/P stocks seems to be choosing those where the profitability or investment premium is not realised in year one. Results confirm that excess returns of the V/P strategy vary due to differences in their market beta, size and book-to-market ratio, with differences in operating profit and investment, only contributing marginally in explaining excess returns of portfolios Q2, Q3, and Q4.\footnote{The effect of the HML falls slightly across the V/P portfolios when the investment and profitability factors are added. This is also consistent with Aharoni et al. (2013) who find a reduction in the coefficient on book-to-market (B/M) ratio when expected investment (growth in equity) is added as an explanatory variable of excess returns. B/M and expected investment are potentially driven by similar economic forces and an improvement in one can be at the expense of the other.}  The alphas from the five-factor model are not significantly different from zero across the quintile V/P portfolios apart from portfolio Q3 which produces a significant negative alpha. Looking at the overlapping returns, a similar pattern is observed on the effect risk factors have on monthly excess returns compared to first-year returns. However, holding portfolios for more than 12 months (overlapping returns) generates a significant positive alpha (0.004) for the highest V/P stocks. This interprets to a $4.9\%$ annual return after risk and demonstrates the impact of extending the portfolio holding period (beyond year one) on monthly excess returns.

By comparing the GRS F-statistics of the previous three models, the five-factor model performs better than either the CAPM or the traditional Fama-French three-factor model. The performance of the four-factor model (unreported), which excludes the B/M factor, is very similar to the five-factor model.\footnote{Fama and French (2015) display that the value factor of the FF three-factor model becomes redundant for describing average return with the addition of profitability and investment factors (GRS F-statistic of the four-factor model is higher than the five-factor one). Excluding the HML factor from the five-factor model, the first-year return results (unreported) reveal that the investment factor becomes positive and significant for the highest V/P portfolio, while, in the case of overlapping returns, both profitability and investment are positive and significant for the same portfolio. Four-factor model alphas are not significant for first year returns but significant positive for the overlapping ones.}  Overall, value-to-price excess returns vary due to differences in market betas, size, book-to-market ratio, operating profit and investment across quintile portfolios and are consistent with Hou et al. (2015) and Fame and French (2015). However, the loadings on the five factors are not evenly increasing from low to high V/P stocks. Finally, profitability and investment provide additional explanatory power across value-to-price portfolio returns but, importantly, the five factors cannot explain all variation in excess returns, especially, for high V/P stocks.

\subsubsection{The effect of second- and third-year returns}
In the previous section, overlapping returns performed significantly better than first year returns after risk. This means that holding stocks that were selected based on the value-to-price ranking of the previous two years contributes significant excess returns to the V/P portfolios. Therefore, interest is turned on the performance of value-to-price portfolios during the second- and third-year holding periods. As reported in Table 6, second- and third-year returns load positively on the market premium and a similar U-shape pattern to the first-year returns is observed. The CAPM alpha is strongly positive (at $1\%$) for the highest V/P stocks and only weekly positive (at $10\%$) for the second- and third highest V/P stocks (Q3, Q4). Third year returns generate a significant positive alpha for the highest and lowest V/P portfolios (0.007 vs. 0.003). On the three-factor model, market, size, and B/M factors have a significant positive effect on the second- and third-year returns with only the size coefficient showing a clear increasing pattern from low to high V/P stocks. For second year returns, significant alphas are observed for the highest V/P stocks (0.005), while, for the third-year returns, alphas are significantly positive for both the highest and lowest V/P stocks (0.006 vs. 0.002). In other words, the highest V/P stocks generate an annual excess return after risk of $6.2\%$ and $7.4\%$ in years two and three, respectively.

Regressing second- and third-year returns on the five factors produces positive and significant loadings with regards to market, size, and B/M factors. Second year returns are positively related to the profitability factor for portfolios with V/P ratio close to one (Q2, Q3, Q4), while this effect although weak extends to the highest V/P stocks when third year returns are considered. The investment factor loads positively on second year returns for portfolios Q3, Q4 and Q5 but its (positive) impact on third year returns seems to concentrate on portfolios Q1, Q2 and Q3. The highest V/P portfolio loads positively on the investment and profitability factors during the second and third year, respectively, a result different to first year and overlapping returns. The investment and profitability factors have a significant positive impact on the lowest and highest V/P portfolios, respectively, despite having no relation to first year and overlapping returns.\footnote{The loading on the value factor is significant negative for the lowest V/P stocks for first year returns. Low V/P stocks resemble growth firms in terms of low book-to-markets and negative HML loadings. The lowest V/P stocks do not load on HML during year two and load positively on HML during year three. However, considering the five-factor model, the lowest of the V/P stocks do not load on the value and profitability factors during years two and three, while the loading on the investment factor is weak negative and strong positive in years two and three, respectively. The negative loading on the investment factor is consistent with the negative effect of expected investment on returns especially for high growth (and potential low V/P) stocks (Aharoni et el.,2013). The performance of the low V/P stocks during year three seems to resemble that of conservative rather than aggressive investment stocks. This is also consistent with the behaviour of growth firms seeking to finance investment initially through equity and later via debt.}  Finally, the alphas of the second and third-year returns are significantly positive only for the highest V/P stocks indicating that holding periods beyond year one produce significant risk-adjusted returns. For example, monthly alphas of 0.004 and 0.005 are generated by the highest V/P stocks in years two and three respectively, and after accounting for all five risk factors proposed in Fama and French (2015). Notably, this study exhibits that a portfolio of stocks with the highest V/P ratio selects companies that are significantly mispriced relative to their equity and profitability growth persistence in the future. For example, the highest V/P stocks offer a significant risk adjusted (excess) return despite loading positively on the investment factor in year two and on the profitability factor in year three. Finally, the second- and third-year performance of the highest V/P stocks explains the difference between the first year and overlapping returns.

\begin{center}
\textbf{[Insert Table 6]}
\end{center}

\subsubsection{The effect of dividends}

Chen et al. (2008) show that the expected value premium, defined as the sum of the difference in the expected dividend price ratio and the difference in the expected long-run dividend growth rate between value and growth portfolios, is on average $6.1\%$ per annum. This premium consists of an expected dividend growth component of $4.4\%$ and an expected dividend price ratio component of $1.7\%$. Thus, a major portion of the value premium comes from the dividend growth component. In the case of overlapping returns, stocks are held for three years, and so, value-to-price portfolio returns are adjusted for dividends. Results for first, second and third-year dividend-adjusted returns are reported in Table 7.
Empirical findings reported in panel A of Table 7 show that the coefficients on the market risk premium are positive and significant for all V/P portfolios. The loadings on the market factor follow the same U-shape that was observed before adjusting portfolio returns for dividends. For first year returns, the alphas are positive and significant (at $5\%$ and $10\%$) with the highest alpha noted among the highest of V/P stocks. Accounting for dividends produced significant and positive alphas for first-year returns, while no significant alphas were observed before adjusting for dividends. In the case of overlapping returns, the V/P strategy yields positive and strongly significant alphas across all V/P portfolios. For example, the highest V/P stocks generate an alpha of 0.007, while the lowest V/P ones produce an alpha of 0.003. For the overlapping returns, CAPM alphas are now significant across all V/P portfolios; however, in the case of non-dividend adjusted returns, alpha was significant only for the highest V/P stocks.

\begin{center}
\textbf{[Insert Table 7]}
\end{center}

Looking at the three-factor model (panel B of Table 7), the coefficients on the market, size and B/M factors are positive and significant for all V/P portfolios. Results confirm that excess returns of the V/P strategy vary due to the differences in size, book-to-market ratio, and market betas across quintile portfolios. Specifically, the highest of the V/P stocks always load higher on size compared to the lowest V/P ones, while their loading on book-to-market is only the third highest. Similar factor loadings are obtained for the overlapping returns. For first year returns, the three-factor model alphas are positive and significant (at $5\%$) only among the highest and lowest value-to-price stocks. In the case of overlapping returns, a positive and significant alpha across all V/P portfolios is observed; the largest alpha (0.005) evidenced for the highest V/P portfolio. The highly significant alphas reported across the V/P portfolios indicate that a risk adjusted return as high as $6.5\%$ could be earned annually. The dividends significantly improved the alphas for the overlapping returns, while, for first year returns, positive (and significant) alphas for the extreme ends of value-to-price stocks were obtained. Recall here, that in the case of simple returns (unadjusted for dividends) alphas from the three-factor model were insignificant.

Panel C of Table 7 report the intercepts and slopes for the five-factor models. For first year returns, market, size, and value factors are positively related to V/P portfolio returns, while loadings demonstrate an increasing (but not monotonic) trend from low to high V/P stock portfolios. Moreover, the coefficients on the profitability (RMW) and investment (CMA) factors are positive and significant only for the V/P portfolios in the middle (Q2, Q3, Q4). Profitability and investment factors do not seem to explain excess returns of the highest and lowest V/P portfolios both for first-year and overlapping returns. The alphas from the five-factor model are positive and weakly significant for portfolios Q1 (low V/P) and Q5 (high V/P). For overlapping returns, a similar pattern is observed about the effect of risk factors and generated alphas. However, alphas become strongly significant with the highest (lowest) V/P portfolio producing a monthly excess return of $0.5\%$ ($0.2\%$). Excluding the HML factor from the five-factor model, empirical results (unreported) reveal that both profitability and investment are positive and significant for the highest V/P portfolio, while alphas remain strongly significant. Finally, adjusting returns for dividends generates positive and strongly (weakly) significant alphas in the case of overlapping (first-year) returns, while, in the case of simple returns (unadjusted for dividends), significant alphas are evidenced only for overlapping returns.

Adjusting second and third-year returns for dividends, produces highly significant alphas across all V/P portfolios for the market and the three-factor model (Table 8). The factor loadings also remain the same.\footnote{This means that the size (market) factor displays an increasing (U-shape) pattern from low to high V/P stocks during both years, while the B/M factor has the second (third) highest effect on the second (third) year returns of the highest V/P stocks.}  Further, high V/P stocks produce significantly higher (risk-adjusted) returns than the low V/P ones in both years. However, in the case of the five-factor model, alphas are only significant for the two extreme V/P portfolios. For example, low V/P stocks yield monthly excess return of $0.2\%$ ($0.3\%$) compared to the $0.5\%$ ($0.6\%$) earned by the high V/P ones, during the second (third) year. In the five-factor model case, market, size, and value risk factors have the same significant positive impact on monthly excess returns as above. The same is true for the profitability coefficients which are significant for portfolios Q2, Q3 and Q4. Importantly, the highest of the V/P stocks load positively on the investment risk factor during year two, while, in year three, it is the lowest of the V/P stocks that load positively on the same factor. Finally, adjusting returns for dividends is producing significant risk-adjusted (excess) returns during years two and three, especially for high V/P stocks, and reinforces the finding that second- and third-year returns contribute significantly on the overlapping returns.

\begin{center}
\textbf{[Insert Table 8]}
\end{center}

Overall, forming portfolios based on value-to-price ratio and holding them for more than a year, produces significant excess returns (after risk) in years two and three. Portfolios generate significant alphas when the holding period extends to longer than a year, with the third-year alphas being slightly higher that the second-year ones and among stocks with the highest value-to-price ratio. Second- and third-year dividend adjusted returns display a similar pattern on the effect of risk factors, while alphas are slightly higher and more significant than the case of simple returns.

\section{Conclusion}
The superior performance of value investing strategies is a well-established empirical fact (Lakonishok et al., 1994; Asness et al., 2013). Value strategies are investments on stocks that appear cheap or have low market prices relative to fundamentals such as earnings, dividends, and book values. The finance literature is mainly using book-to-market ratios to identify value stocks, while the accounting one introduced a value-to-price ratio, where the intrinsic value of the firm is estimated using the residual income valuation model (Frankel and Lee, 1998; Ali et al., 2003; Goncalves and Leonard, 2023; Cong et al., 2023). Frankel and Lee (1998) find that a value-to-price (V/P) strategy is more successful and leads to abnormal returns, at longer horizons, than simple market-multiples do.\footnote{One explanation of this slow price convergence is the speed at which long-term fundamental information is incorporated in stock prices. An alternative explanation of the value-to-price effect is that it reflects cross-sectional risk differences.}  Frankel and Lee (1998) and Ali et al., (2003) claim that the predictive ability of V/P strategy is most probably due to market mispricing. On the contrary, Hwang and Lee (2013) suggest that the mispricing explanation of the V/P anomaly is over-hasty and further research is necessary. Motivated by findings above, this paper investigates, first, whether investments in high value-to-price stocks outperform investments in low value-to-price stocks, second, the relationship between V/P ratios and other risk proxies at firm level and, third, whether asset pricing factor models can explain excess return of  V/P portfolios (Hou et al., 2015; Fama and French, 2015). Overall, the aim of this study is to examine the risk vs mispricing explanation of value investing using value-to-price ratio sorted portfolios. To answer these questions, data from the merger of COMPUSTAT, CRSP, I/B/E/S is used for all non-financial firms listed in AMEX, NYSE, and NASDAQ in the period from 1987 to 2015.

Results show that high value-to-price portfolios outperform low value-to-price portfolios. A hedge portfolio formed by the V/P ratio produced on average raw (size adjusted) buy-and-hold returns of $5.3\%$ ($3.2\%$), $13.7\%$ ($8.8\%$) and $27.8\%$ ($14.5\%$) over the next 12 months, 24 months, and 36 months, respectively. The predictive power of the V/P strategy is comparable to that of the B/M strategy in the short term (with a one-year horizon). However, the performance of the V/P strategy significantly improved over longer horizons in comparison with those of the B/M. For instance, the performance of the V/P hedge portfolio spread over 36 months was $27.8\%$ ($14.5\%$), compared with only $11.1\%$ ($10.6\%$) for the B/M hedge portfolios. Interestingly, portfolios with the highest value-to-price ratio also have the lowest market value of equity but the size effect explains only part of the difference in returns and primarily in the first year. Overall, the V/P effect reported here was highly consistent with that reported in other studies such as Frankel and Lee (1998), Ali et al. (2003), Goncalves and Leonard (2023), Cong et al., (2023).

To investigate the risk explanation of V/P strategies, the relationship between the V/P ratio and firm characteristics (market beta, size, book-to-market ratio, idiosyncratic volatility, earnings variability, leverage, bankruptcy and analyst coverage), which are known to be proxies for risk, is examined. Results also highlight a positive and significant association between the value-to-price ratio and idiosyncratic volatility, earnings variability, and leverage, and a negative one between value-to-price ratio and size. In other words, firms with a high V/P ratio are riskier and potentially require higher expected returns. In addition, this study examined the relationship between the V/P ratio and long horizon (buy-and-hold) stock returns after controlling for the previous risk factors. Results confirm that the coefficient on the V/P remains significant and positive after controlling for various risk characteristics; thus, confirming that omission of risk proxies is not an explanation of the V/P effect.

Furthermore, the study investigates the ability of Fama and French's five-factor model to explain V/P strategy excess returns. For first year returns, the coefficients on the risk factors are positive and significant, while alphas are not significantly different from zero. Specifically, the market, size and value factors are positively related to V/P portfolio excess returns with loadings showing an increasing, but not monotonic, pattern shifting from low to high V/P stocks. The coefficients on the profitability (RMW) and investment (CMA) factors are positive and significant only for the V/P portfolios in the middle (Q2, Q3, Q4). Interestingly, the profitability and investment factors do not explain excess returns of the highest and lowest V/P portfolios. In Novy-Marx (2013) stocks with high B/M and high gross profit also realize high stock returns. Given that the V/P ratio is equivalent to adding the present value of future abnormal profits on the nominator of the B/M ratio, a portfolio of stocks with the highest V/P ratio consists of stocks where the profitability or investment premium is not realised in year one. For overlapping returns, risk factor loadings are like first-year ones, while significant alphas are observed only for portfolios with the highest V/P stocks. Holding portfolios for more than 12 months (overlapping returns) generates an annual return of $4.9\%$ after risk for the highest V/P stocks and raises questions about the importance of second- and third-year returns. Forming portfolios on the basis of value-to-price ratio and holding them for more than a year, produces significant excess returns (after risk) in years two and three, with the third-year alphas being slightly higher that the second-year ones and mainly among the highest of the V/P stocks.\footnote{For example, the highest V/P stocks offer a significant risk adjusted (excess) return despite loading positively on the investment factor in year two and on the profitability factor in year three. More, the second- and third-year performance of the highest V/P stocks explains the difference between the first year and overlapping returns. Adjusting second- and third-year returns for dividends, a similar pattern on the effect of the risk factors is obtained, while the observed alphas are slightly higher and more significant than the case of simple returns, across all years and V/P portfolios.}

Finally, results indicate that V/P returns are largely explained by exposure to market, size, value, investment, and profitability risk factors. However, these factors cannot explain all variation in excess returns and the V/P ratio, while portfolios with the highest V/P stocks select companies that are significantly mispriced relative to their equity (investment) and profitability growth persistence in the future.

\clearpage

\section*{References}
Ali, A., Hwang, L. and Trombley, M.A., (2003) Residual-income-based valuation predicts future stock returns: Evidence on mispricing vs. risk explanations. The Accounting Review, 78(2), 377-396.

\noindent
Aharoni, G., Gundy, B., and Zeng, Q., (2013) Stock returns and the Miller Modigliani valuation formula: Revisiting the Fama French analysis. Journal of Financial Economics, 110, 347-357.

\noindent
Altman, E.I., (1968) Financial ratios, discriminant analysis and the prediction of corporate bankruptcy. The Journal of Finance, 23(4), 589-609.

\noindent
Ashton, D., and Wang, P., (2015) Conservatism in residual income models: theory and supporting evidence. Accounting and Business Research, 45(3), 387-410.

\noindent
Asness, C.S., Moskowitz, T.J., and Pedersen, L.H., (2013) Value and momentum everywhere. The Journal of Finance, 68, 929-985.

\noindent
Avery, R.B., (1977) Error components and seemingly unrelated regressions. Econometrica, 45, 199-209.

\noindent
Ball, R., Gerakos, J., Linnainmaa, T.J., and Nikolaev, V., (2016) Accruals, cashflows, and operating profitability in the cross section of stock returns. Journal of Financial Economics, 121, 28-45.

\noindent
Ball, R., Gerakos, J., Linnainmaa, T.J., and Nikolaev, V., (2020) Earnings, retained earnings, and book-to-market in the cross section of expected returns. Journal of Financial Economics, 135, 231-254.

\noindent
Baltagi, B.H., (1980) On seemingly unrelated regressions with error components. Econometrica, 48, 1547-1551.

\noindent
Barillas, F., and Shanken, J., (2018) Comparing asset pricing models. The Journal of Finance, 73(2), 715-754.

\noindent
Barth, M.E., Beaver, W.H., Hand, J.R. and Landsman, W.R., (1999) Accruals, cash flows, and equity values. Review of Accounting Studies, 4(3-4), 205-229.

\noindent
Barth, M.E., Beaver, W.H., Hand, J.R. and Landsman, W.R., (2005) Accruals, accounting-based valuation models, and the prediction of equity values. Journal of Accounting, Auditing $\&$ Finance, 20(4), 311-345.

\noindent
Barth, M.E. and Hutton, A.P., (2004) Analyst earnings forecast revisions and the pricing of accruals. Review of Accounting Studies, 9, 59-96.
Beaver, W.H. (2002) Perspectives on recent capital market research. The Accounting Review, 77(2), 453-474.

\noindent
Biorn, E., (2004) Regression systems for unbalanced panel data: A stepwise maximum likelihood procedure. Journal of Econometrics, 122, 281-291.

\noindent
Brennan, M.J., Jegadeesh, N. and Swaminathan, B., (1993) Investment analysis and the adjustment of stock prices to common information. Review of Financial Studies, 6(4), 799-824.

\noindent
Brennan, M.J., and Subrahmanyam, A., (1995) Investment analysis and price formation in securities markets. Journal of Financial Economics, 38(3), 361-381.

\noindent
Bryan, D.M. and Tiras, S.L., (2007) The influence of forecast dispersion on the incremental explanatory power of earnings, book value, and analyst forecasts on market prices. The Accounting Review, 82(3), 651-677.

\noindent
Chen, L., Petkova, R. and Zhang, L., (2008) The expected value premium. Journal of Financial Economics, 87(2), 269-280.

\noindent
Cong, L.W., George, N.D., and Wang, G., (2023) RIM-based value premium and factor pricing using value-price divergence. Journal of Banking and Finance, 149, 1-22.

\noindent
Daniel, K. and Titman, S., (1997) Evidence on the characteristics of cross-sectional variation in stock returns. The Journal of Finance, 52(1), 1-33.

\noindent
Dechow, P.M., and Sloan, R.G., (1997) Returns to contrarian investment strategies: Tests of naive expectations hypotheses. Journal of Financial Economics, 43, 3-27.

\noindent
Dechow, P.M., Hutton, A.P., and Sloan, R.G., (1999) An empirical assessment of the residual income valuation model. Journal of Accounting and Economics, 26, 1-34.

\noindent
DeBondt, W.F.M. and Thaler, R.H., (1985) Does the stock market overreact? The Journal of Finance, 40, 793-805.

\noindent
Fama, E.F. and French, K.R., (2015) A five-factor asset pricing model. Journal of Financial Economics, 116(1), 1-22.

\noindent
Fama, E.F. and French, K.R., (2016) Dissecting anomalies with a five-factor model. Review of Financial Studies, 29(1), 69-103.

\noindent
Fama, E.F. and French, K.R., (2006) Profitability, investment, and average returns. Journal of Financial Economics, 82(3), 491-518.

\noindent
Fama, E.F. and French, K.R., (1997) Industry costs of equity. Journal of Financial Economics, 43(2), 153-193.

\noindent
Fama, E.F. and French, K.R., (1995) Size and book-to-market factors in earnings and returns. The Journal of Finance, 50(1), 131-155.

\noindent
Fama, E.F. and French, K.R., (1993) Common risk factors in the returns on stocks and bonds. Journal of Financial Economics, 33(1), 3-56.

\noindent
Fama, E.F. and French, K.R. (1992) The cross-section of expected stock returns. The Journal of Finance, 47(2), 427-465.

\noindent
Feltham, G.A. and Ohlson, J.A., (1995) Valuation and clean surplus accounting for operating and financial activities. Contemporary Accounting Research, 11(2), 689-731.

\noindent
Frankel, R. and Lee, C.M., (1998) Accounting valuation, market expectation, and cross-sectional stock returns. Journal of Accounting and Economics, 25(3), 283-319.

\noindent
Gallant, R.A., (1975)Seemingly unrelated nonlinear regressions. Journal of Econometrics, 3, 35-50.

\noindent
Gebhardt, W.R., Lee, C., and Swaminathan, B. (2001) Toward an implied cost of capital. Journal of Accounting Research, 39(1), 135-176.

\noindent
Gibbons, M.R., Ross, S.A., and Shanken, J., (1989) A test of the efficiency of a given portfolio. Econometrica, 57, 1121-1152.

\noindent
Gode, D., and Mohanram, P., (2003) Inferring the cost of capital using the Ohlson-Juettner model. Review of Accounting Studies, 8(4), pp. 399-431.

\noindent
Golubov, A., and Konstantinidi, T., (2019) Where is the risk in value? Evidence from a market-to-book decomposition. The Journal of Finance, 74(6), 3135-3186.

\noindent
Goncalves, S.A., and Leonard, G., (2023) The fundamental-to-market ratio and the value premium decline. Journal of Financial Economics, 147, 382-405.

\noindent
Gregory, A., Harris, R.D.F., and Michou, M., (2001) An analysis of contrarian investment strategies in the UK. Journal of Business Finance and Accounting, 28(9), 1131-1228.

\noindent
Gregory, A., Harris, R.D.F., and Michou, M., (2003) Contrarian investment and macroeconomic risk. Journal of Business Finance and Accounting, 30(1), 213-255.

\noindent
Hahn, J., and Lee, H., (2006) Yield spreads as alternative risk factors for size and book-to-market. Journal of Financial and Quantitative Analysis, 41(2), 245-269.

\noindent
Hou, K., Xue, C., and Zhang, L., (2015) Digesting anomalies: An investment approach. Review of Financial Studies, 28(3), 650-705.

\noindent
Hwang, L. and Lee, W., (2013) Stock return predictability of residual-income-based valuation: Risk or mispricing? Abacus, 49(2), 219-241.

\noindent
Jaffe, F.J., Jindra, J., Pedersen, J.D., and Voetmann, T., (2019) Can mispricing explain the value premium? Financial Management, 49, 615-633.

\noindent
Kelly, B. T., Pruitt, S., and Su, Y., (2019) Characteristics are covariances: A unified model of risk and return. Journal of Financial Economics, 134, 501-524.

\noindent
Kinal, T., and Lahiri, K., (1990) A computational algorithm for multiple equation models with panel data. Economics Letters, 34, 143-146.

\noindent
Kothari, S., (2001) Capital markets research in accounting. Journal of Accounting and Economics, 31(1), 105-231.

\noindent
Lakonishok, J., Shleifer, A. and Vishny, R.W., (1994) Contrarian investment, extrapolation, and risk. The Journal of Finance, 49(5), 1541-1578.

\noindent
Lee, C.M. and Swaminathan, B., (1999) Valuing the Dow: A bottom-up approach. Financial Analysts Journal, 55(5), 4-23.

\noindent
Lee, C., Myers, J. and Swaminathan, B., (1999) What is the intrinsic value of the Dow? The Journal of Finance, 54(5), pp. 1693-1741.

\noindent
Lettau, M., and Ludvigson, S., (2001) Resurrecting the (C)CAPM: A Cross-sectional test when risk premia are time-varying. Journal of Political Economy, 109, 1238-1287.

\noindent
Lo, K. and Lys, T., (2000) The Ohlson model: contribution to valuation theory, limitations, and empirical applications. Journal of Accounting, Auditing $\&$ Finance, 15(3), 337-367.

\noindent
McElroy, M.B., and Burmeister, E., (1988) Arbitrage pricing theory as a restricted nonlinear multivariate regression model. Journal of Business and Economic Statistics, 6(1), 29-42.

\noindent
Myers, J.N., (1999) Implementing residual income valuation with linear information dynamics. The Accounting Review, 74(1), 1-28.

\noindent
Newey, W.K. and West, K.D., (1987) Hypothesis testing with efficient method of moments estimation. International Economic Review, 777-787.

\noindent
Novy-Marx, R., (2013) The other side of value: The gross profitability premium. Journal of Financial Economics, 108, 1-28.

\noindent
Ohlson, J.A., (2001) Earnings, book values, and dividends in equity valuation: An empirical perspective. Contemporary Accounting Research, 18(1), 107-120.

\noindent
Ohlson, J.A., (1990) A synthesis of security valuation theory and the role of dividends, cash flows, and earnings. Contemporary Accounting Research, 6, 648-676.

\noindent
Ohlson, J.A., (1995) Earnings, book values and dividends in security valuation. Contemporary Accounting Research, 11, 661-687.

\noindent
Peasnell, K.V., (1982) Some formal connections between economic values and yields and accounting numbers. Journal of Business Finance $\&$ Accounting, 9(3), 361-381.

\noindent
Penaranda, F., and Sentana, E., (2015) A unifying approach to the empirical evaluation of asset pricing models. The Review of Economics and Statistics, 97(2), 412-435.

\noindent
Petkova, R., (2006) Do the Fama-French factors proxy for innovations in predictive variables. The Journal of Finance, 61(2), 581-612.

\noindent
Petkova, R., and Zhang, L., (2005) Is value riskier than growth? Journal of Financial Economics, 78, 187-202.

\noindent
Piotroski, J.D., and So, E.C., (2012) Identifying expectation errors in value/glamour strategies: A fundamental analysis approach. Review of Financial Studies, 25, 2841-2875.

\noindent
Pope, P.F. and Wang, P., (2005) Earnings components, accounting bias and equity valuation. Review of Accounting Studies, 10(4), 387-407.

\noindent
Sarafidis, V., and T. Wansbeek (2012) Cross-Sectional Dependence in Panel Data Analysis. Econometric Reviews, 31(5), 483-531.

\noindent
Sarafidis, V., and T. Wansbeek (2021) 40 years of panel data analysis: past, present and future. Journal of Econometrics, 220(1), 215-226.

\noindent
Sentana, E., (2009) The econometrics of mean-variance efficiency tests: a survey, Econometrics Journal, 12, 65-101.

\noindent
Shleifer, A. and Vishny, R.W., (1997) The limits of arbitrage. The Journal of Finance, 52(1), 35-55.

\noindent
Tsay, R.S., Lin, Y. and Wang, H., (2008) Residual income, value-relevant information, and equity valuation: A simultaneous equations approach. Review of Quantitative Finance and Accounting, 31(4), 331-358.

\noindent
Tsay, R.S., Lin, Y. and Wang, H., (2009) Residual income, non-earnings information, and information content. Journal of Forecasting, 28(6), 487-511.

\noindent
Wang, P., (2013) The role of disaggregation of earnings in stock valuation and earnings forecasting. Accounting and Business Research, 43(5), 530-557.

\noindent
Wei, J.K. and Zhang, J., (2007) Arbitrage risk and arbitrage returns: Evidence from the fundamental value-to-price anomaly. Working paper, Hong Kong University of Science and Technology.

\noindent
Johnson, W. B., and Xie, S., (2004) The convergence of stock price to fundamental value. Social Science Research Network, Elsevier.

\noindent
Xu, L., (2007) Is V/P a distinct anomaly? Review of Accounting and Finance, 6(4), 404-418.

\noindent
Zellner, A., (1962) An efficient method of estimating seemingly unrelated regressions and tests for aggregation bias. Journal of the American Statistical Association, 57, 348-368.

\clearpage

\section*{Appendix}
\subsection*{Theoretical development of the residual income valuation model}

The most popular accounting-based approach used to predict firms, value is the Residual Income Model (RIM, hereafter), developed by Peasnell (1982), Ohlson (1990, 1995), and Feltham and Ohlson (1995). According to the RIM model, the value of any firm can be expressed as a function of its current year book value plus the present value of expected future residual income\footnote{Residual income is defined as the difference between the investors, expected income and the required income, where the required income is calculated as forecast book equity at the start of each period multiplied by the cost of equity capital.}, as shown below:

\begin{align}
MV_t = BV_t + \sum_{T=1}^{\infty} R^{-T} \, \mathbb{E}_t \left(NI_{t+T}^a\right), \tag{A1}
\end{align}
where $MV_t$ is the market value of equity at date $t$; $BV_t$ is the book value of equity at date $t$; $R = 1 + r$ (with $r$ being the cost of equity capital); $\mathbb{E}_t[\cdot]$ is the expectation operator based on information available at time $t$; $NI_t$ is the net income for period $t$; and $NI_t^a$ is the residual income defined as $NI_t - r \times BV_{t-1}$.

Frankel and Lee (1998) use the residual income model by simplifying the valuation technique over a short horizon and assuming that, after the third year, the forecasted residual income will be earned in perpetuity.\footnote{Thus, the market value calculation is based on a three-year horizon, as shown below:
\begin{align*}
MV_t = BV_t
+ \frac{(FROE_t - r)}{(1 + r)} BV_t
+ \frac{(FROE_{t+1} - r)}{(1 + r)^2} BV_{t+1}
+ \frac{(FROE_{t+2} - r)}{(1 + r)^2 r} BV_{t+2},
\end{align*}
where $BV_{t+k}$ is the forecast book value of equity at the end of year $t+k$, $k = 1, 2,$ or $3$; $FROE_{t+k}$ is the forecast return on equity for year $t+k$, $k = 1, 2,$ or $3$; and $r$ is the estimated cost of equity capital.}

Their empirical implementation of the residual income model requires estimating the future book value of equity and the future return on equity for the next three years.\footnote{Two alternative approaches to estimate $FROE_t$ are used. The first approach is based on the earnings in the previous period, while the second is based on I/B/E/S analysts' forecasts. To estimate the former, the authors use the return on equity for period $t$ to proxy for all three future returns on equity ($FROE_{t+k}$). To estimate the latter, they use one-year-ahead and two-year-ahead consensus I/B/E/S earnings-per-share forecasts, as well as a five-year long-term earnings growth rate in earnings.}

Frankel and Lee,s (1998) implementation of the residual income model differs from the original implementation by Ohlson (1995) and Dechow et al.\ (1999). Ohlson (1995) argues that a firm,s ability to generate residual income is driven by its monopolistic power. However, this monopolistic power will diminish over time due to market competition, residual income will shrink, and the returns earned by the firm will eventually equal the cost of capital. To capture this process, an autoregressive process can be used to model and forecast abnormal earnings.

According to Ohlson,s linear information dynamics, the abnormal income in period $t+1$ is a linear function of the abnormal income in the current period and other information ($\nu$).\footnote{Other information represents any relevant information other than accounting information. According to Ohlson (1995), other information in the next period ($t+1$) is a linear function of other information from the current period ($t$).} Thus, Ohlson,s linear information model can be expressed using Equation~(A2):

\begin{align}
NI_{it}^a &= \omega NI_{it-1}^a + \nu_{it-1} + \varepsilon_{1,it} \tag{A2$\alpha$} \\
\nu_{it} &= \gamma \nu_{it-1} + \varepsilon_{2,it} \tag{A2b} \\
MV_{it} &= BV_{it} + \alpha_1 NI_{it}^a + \alpha_2 \nu_{it} + u_{it} \tag{A2c}
\end{align}

\noindent where $\nu_{it}$ is other information; $\varepsilon_{1,it}$, $\varepsilon_{2,it}$ and $u_{it}$ are error terms; the subscripts $i$ and $t$ refer to firm and year, respectively; and $\omega$ and $\gamma$ are persistence parameters such that $0 \leq \omega < 1$ and $0 \leq \gamma < 1$.

\begin{align*}
\alpha_1 &= \frac{\omega}{1 + r - \omega}; \quad
\alpha_2 = \frac{1 + r}{(1 + r - \omega)(1 + r - \gamma)}.
\end{align*}
Feltham and Ohlson (1995) argue neither that accounting measures of performance are neutral nor that competitive power will, over time, drive residual income to zero, as proposed by Ohlson (1995). On the contrary, accounting practices $-$particularly accounting conservatism$-$ cause the book value of equity to differ systematically from the market value of equity. In other words, accounting conservatism influences the residual income series in the long run, because it understates the book value of equity.\footnote{Book value of equity is used as a benchmark to calculate normal returns and, consequently, residual income, as shown in Equation A1.}

Thus, Feltham and Ohlson (1995) suggest a second linear information process in which the book value of equity is used as a proxy for conservatism. Myers (1999) emphasizes that the key contribution of Ohlson (1995) and Feltham and Ohlson (1995) stems from their linear information dynamics. He argues that ad hoc modifications of these linear information dynamics, as in Frankel and Lee (1998) and Dechow et al.\ (1999), could violate the internal consistency of the model.\footnote{Myers (1999) maintains that intrinsic value calculation as implemented by Frankel and Lee in one part of their model, often implies arbitrage. Therefore, he proposes a framework for modifying linear information dynamics while preserving the internal consistency of the model.}

Similarly, Ohlson (2001) contends that ignoring the other information variable ($v$), or equating it to zero as proposed by Dechow et al.\ (1999), may be of empirical interest. However, these propositions drastically reduce the empirical content of the linear information model. More importantly, Ohlson (2001) states that it is plausible to use a consensus of analysts, forecasts ($f_t$) for period $t+1$ as a proxy for expected earnings based on all available information at period $t$, and hence to calculate the other information variable ($v$). According to Dechow et al.\ (1999) and Ohlson (2001), the other information variable ($v$) can be calculated as follows:

\begin{align}
v_t &= \mathbb{E}_t \left[ NI_{t+1}^a \right] - \omega \times NI_t^a; \tag{A3} \\
\mathbb{E}_t \left[ NI_{t+1}^a \right] &= f_t^a = f_t - r \times BV_t; \notag \\
v_t &= f_t^a - \omega \cdot NI_t^a, \notag
\end{align}
where $\mathbb{E}_t \left[ NI_{t+1}^a \right]$ is the conditional expectation of abnormal income for period $t+1$ based on all information available at period $t$; $f_t$ is the consensus of analysts, forecasts of expected earnings for period $t+1$; and $\omega$ is the persistence parameter of abnormal income, estimated by ignoring the other information variable in Equation~(A2$\alpha$).

\subsection*{Risk proxies}
This section outlines the definition, measurement and economic intuition behind each firm characteristic that we use as a proxy for risk.

$Beta:$ This is a measure of systematic risk. Beta is estimated for each firm-year by implementing the Capital Asset Pricing Model (CAPM). Previous studies have documented a positive relationship between a firm's specific Beta and future stock returns (Fama and French, 1992; Frankel and Lee, 1998; Ali et al., 2003). We use the CRSP value-weighted index as a proxy for market returns.

Firm-specific Betas are estimated at the end of December each year by regressing the monthly returns of each firm against the contemporaneous monthly returns of the CRSP value-weighted index, using the previous 36 months of data. In other words, to estimate the Beta of firm $i$ for year $t$, we use firm $i$'s monthly returns over the period from January $t{-}3$ to December $t$.

$Ivolatility$: This is a measure of unsystematic or idiosyncratic risk. Unsystematic risk for each firm-year is calculated as follows. First, we regressed the daily returns data for the previous year at the end of December each year, against the contemporaneous daily returns of the CRSP value weighted index. Second, we used the variance of the residuals from the previous regression as a proxy of Ivolatility. Several previous empirical studies have documented an association between future stocks returns and idiosyncratic risk (Ali et al., 2003; Gebhardt et al., 2001; Gode and Mohanram, 2003).

$D/M$: This is a measure of leverage in the firm. Several prior studies have suggested a positive association between a firm,s future stocks returns and its leverage ratio (Fama and French, 1992; Gebhardt et al., 2001; Gode and Mohanram, 2003). For every year, we measured $D/M$ as the ratio of the book value of long-term debt at the end of December of the previous year to the market capitalization at the end of June in the current year.
$Ln(ME)$: This is a measure of firm size. Several previous studies use firm size as a proxy of the information environment. It is argued that the information environment is influenced by several interrelated factors such as trading volume, bid-ask spreads, firm size and institutional investors (Barth and Hutton, 2004). It is expected that firms with a better information environment have a lower risk premium because it reduces the information asymmetry between the firm and investors (Ali et al., 2003). It is well documented that size is correlated with the differences in information environment that lead to a lower risk for large firms than for small firms (Gebhardt et al., 2001; Gode and Mohanram, 2003). For the purposes of this study, size is measured as the log of firm i,s market value at the end of June in year t. A negative association between the risk premium and firm size is expected.

$Analysts$: They are a measure of the financial analysts, coverage of the firm. It is another measure of the information environment. For instance, Brennan et al. (1993) argue that a firm with better coverage from financial analysts responds faster to market information than those with inferior analysts, coverage. Furthermore, analysts, coverage can be used as a proxy of firm liquidity.

For instance, Brennan and Subrahmanyam (1995) argue that firms with better analysts, coverage tend to be more liquid than those with inferior analysts, coverage. Therefore, we used the number of analysts, estimates included in the I/B/E/S database in May of year t as a proxy for liquidity and information environment. We expected a negative association between firms with better analysts, coverage and future stock returns.

$Altman's$ $Z$: This is a measure of financial distress. It is measured as a bankruptcy score, from Altman,s (1968) model:
\begin{align*}
\text{Altman's } Z = &\ 1.2 \times \left( \frac{\text{Working Capital}}{\text{Total Assets}} \right)
+ 1.4 \times \left( \frac{\text{Retained Earnings}}{\text{Total Assets}} \right) \\
&+ 3.3 \times \left( \frac{\text{EBIT}}{\text{Total Assets}} \right)
+ 0.6 \times \left( \frac{\text{Market Value of Equity}}{\text{Book Value of Total Liabilities}} \right) \\
&+ 1.0 \times \left( \frac{\text{Sales}}{\text{Total Assets}} \right).
\end{align*}
We expect a positive association between Altman,s Z score and future stock returns.

Std(ROA): This is a measure of earnings variability. Several previous studies have argued that the variability of earnings is likely to reflect intrinsic cash flow risks and is considered a main source of risk for firm valuations (Gebhardt et al., 2001; Gode and Mohanram, 2003). For the purposes of this paper, Std(ROA)  is calculated as the standard deviation of returns on assets in the past five years.

$B/M$: This is a measure of the book to market ratio. It has been argued that $B/M$ can be used as a proxy for accounting conservatism, the growth opportunities of a firm, or perceived risk (Fama and French, 1992; Lakonishok et al., 1994). It is very difficult to distinguish empirically between various interpretations of B/M and to predict the direction of the relationship between the B/M ratio and future stock returns. For the purposes of this paper, it is calculated as the book value of equity at the end of December of the previous year divided by the market value of equity at the end of June in the current year.

\newpage
\newgeometry{left=.5in, right=1in, top=0.5in, bottom=0.5in}
\captionsetup{labelformat=empty}
\begin{table}[htbp]
\centering
\caption*{\textbf{Table A1:} Number of observations by year and industry (1987--2000)}
\begin{tabular}{lrrrrrrrrrrrr}
\toprule
\toprule
Year & Non-Dur. & Dur. & Manufac. & Energy & Chem. & Equip. & Tel. & Util. & Retail & Health &  Other & Total\\
\midrule
\midrule
1987 & 22 & 12 & 41 & 14 & 14 & 31 & 5 & 34 & 12 & 15 & 41 & 241\\
1988 & 23 & 12 & 41 & 14 & 13 & 33 & 5 & 37 & 13 & 18 & 46 & 255\\
1989 & 23 & 12 & 43 & 19 & 14 & 35 & 5 & 37 & 13 & 20 & 48 & 269\\
1990 & 23 & 10 & 46 & 18 & 17 & 37 & 7 & 42 & 14 & 23 & 48 & 285\\
1991 & 23 & 10 & 54 & 21 & 17 & 37 & 8 & 43 & 15 & 25 & 49 & 302\\
1992 & 25 & 13 & 57 & 21 & 18 & 44 & 8 & 43 & 19 & 30 & 50 & 328\\
1993 & 26 & 16 & 72 & 28 & 17 & 50 & 12 & 45 & 22 & 32 & 62 & 382\\
1994 & 27 & 18 & 81 & 30 & 21 & 53 & 13 & 45 & 27 & 36 & 70 & 421\\
1995 & 29 & 19 & 95 & 35 & 23 & 74 & 14 & 48 & 29 & 38 & 77 & 481\\
1996 & 34 & 19 & 100 & 39 & 23 & 89 & 18 & 52 & 37 & 48 & 93 & 552\\
1997 & 36 & 21 & 107 & 46 & 24 & 99 & 18 & 57 & 38 & 48 & 112 & 606\\
1998 & 35 & 21 & 108 & 42 & 25 & 113 & 18 & 57 & 44 & 55 & 117 & 635\\
1999 & 38 & 21 & 109 & 51 & 26 & 129 & 23 & 59 & 47 & 59 & 127 & 689\\
2000 & 36 & 21 & 108 & 58 & 27 & 139 & 22 & 63 & 44 & 66 & 122 & 706\\
2001 & 43 & 20 & 111 & 64 & 25 & 114 & 25 & 64 & 47 & 67 & 122 & 702\\
2002 & 46 & 21 & 114 & 66 & 28 & 133 & 31 & 65 & 52 & 72 & 150 & 778\\
2003 & 49 & 24 & 125 & 71 & 29 & 173 & 38 & 66 & 57 & 82 & 158 & 872\\
2004 & 50 & 25 & 136 & 73 & 31 & 182 & 40 & 68 & 71 & 85 & 177 & 938\\
2005 & 51 & 29 & 142 & 86 & 33 & 202 & 47 & 72 & 79 & 84 & 196 & 1021\\
2006 & 54 & 30 & 158 & 99 & 39 & 214 & 50 & 76 & 84 & 94 & 226 & 1124\\
2007 & 57 & 35 & 164 & 108 & 39 & 230 & 54 & 79 & 88 & 100 & 238 & 1192\\
2008 & 57 & 25 & 151 & 98 & 34 & 183 & 47 & 85 & 85 & 101 & 230 & 1096\\
2009 & 63 & 34 & 175 & 109 & 42 & 235 & 50 & 89 & 93 & 115 & 255 & 1260 \\
2010 & 69 & 39 & 194 & 123 & 45 & 280 & 53 & 90 & 101 & 123 & 280 & 1397\\
2011 & 70 & 46 & 186 & 145 & 48 & 284 & 62 & 90 & 104 & 123 & 292 & 1450\\
2012 & 73 & 43 & 192 & 144 & 52 & 304 & 63 & 93 & 113 & 126 & 304 & 1507\\
2013 & 79 & 46 & 205 & 152 & 54 & 330 & 69 & 94 & 141 & 138 & 344 & 1652\\
2014 & 85 & 48 & 197 & 121 & 60 & 350 & 71 & 107 & 141 & 159 & 357 & 1696\\
Total & 1246 & 690 & 3312 & 1895 & 838 & 4177 & 876 & 1800 & 1630 & 1982 & 4391 & 22837\\
\% & 5.46 & 3.02 & 14.5 & 8.3 & 3.67 & 18.29 & 3.84 & 7.88 & 7.14 & 8.68 & 19.23 & 100\\
\bottomrule
\end{tabular}

\vspace{0.5em}
  \begin{minipage}{\textwidth}
    \footnotesize \textit{\textbf{Notes}:} \\
1) \textbf{Non-Dur.} - Consumer Non-Durables - Food, Tobacco, Textiles, Apparel, Leather, Toys (SIC code: 0100--0999, 2000--2399, 2700--2749, 2770--2799, 3100--3199, 3940--3989) \\
2) \textbf{Dur.} - Consumer Durables - Cars, TVs, Furniture, Household Appliances (SIC code: 2500--2519, 2590--2599, 3630--3659, 3710--3711, 3714, 3716, 3750--3751, 3792, 3900--3939, 3990--3999) \\
3) \textbf{Manufac.} - Manufacturing - Machinery, Trucks, Planes, Office Furniture, Paper, Com Printing (SIC code: 2520--2589, 2600--2699, 2750--2769, 3000--3099, 3200--3569, 3580--3629, 3700--3709, 3712--3713, 3715, 3717--3749, 3752--3791, 3793--3799, 3830--3839, 3860--3899) \\
4) \textbf{Energy} - Oil, Gas and Coal Extraction and Products (SIC code: 1200--1399, 2900--2999) \\
5) \textbf{Chemical} - Chemicals and Allied Products (SIC code: 2800--2829, 2840--2899) \\
6) \textbf{Equip.} - Business Equipment, Computers, Software, Electronic Equipment (SIC code: 3570--3579, 3660--3692, 3694--3699, 3810--3829, 7370--7379) \\
7) \textbf{Telecom} - Telephone and Television Transmission (SIC code: 4800--4899) \\
8) \textbf{Utility} - Utilities (SIC code: 4900--4949) \\
9) \textbf{Retail} - Wholesale, Retail, Services (e.g. Laundries, Repair Shops) (SIC code: 5000--5999, 7200--7299, 7600--7699) \\
10) \textbf{Health} - Healthcare, Medical Equipment, Drugs (SIC code: 2830--2839, 3693, 3840--3859, 8000--8099) \\
11) \textbf{Other} - Other Mines, Construction, Materials, Transport, Hotels, Services, Entertainment.
 \end{minipage}
\end{table}
\restoregeometry



\section*{Supplementary material}

Table 1 Characteristics of quantile-portfolio formed by ME, B/P and V/P ratios

Panel A- Market equity portfolios (In sample size quintiles)

|  | Q1<br>Low ME | Q2 | Q3 | Q4 | Q5<br>High ME | All Firms | Q5-Q1<br>Diff. |
|---|---|---|---|---|---|---|---|
| ME | 274 | 1014 | 2392 | 5837 | 20491 | 6984 | 20216*** |
| V/P | 2.243 | 1.285 | 1.081 | 0.992 | 0.904 | 1.476 | -1.339*** |
| B/M | 1.670 | 1.110 | 0.859 | 0.692 | 0.470 | 1.104 | -1.201*** |
| beta | 1.262 | 1.291 | 1.209 | 1.164 | 1.003 | 1.186 | -0.259*** |
| Ret12 | 0.187 | 0.140 | 0.151 | 0.130 | 0.109 | 0.152 | -0.078*** |
| Ret24 | 0.411 | 0.329 | 0.324 | 0.298 | 0.247 | 0.340 | -0.168*** |
| Ret36 | 0.585 | 0.426 | 0.366 | 0.388 | 0.304 | 0.447 | -0.286*** |
| SRet12 | 0.069 | 0.027 | 0.028 | 0.005 | 0.005 | 0.035 | -0.063*** |
| SRet24 | 0.124 | 0.062 | 0.043 | 0.019 | 0.017 | 0.067 | -0.106*** |
| SRet36 | 0.179 | 0.097 | 0.028 | 0.037 | 0.027 | 0.094 | -0.151*** |
| Obs. | 3325 | 3314 | 3315 | 3296 | 3330 | 16,580 | |

Panel B- Book to Market (B/M) portfolios

|  | Q1<br>Low B/M | Q2 | Q3 | Q4 | Q5<br>High B/M | All Firms | Q5-Q1<br>Diff. |
|---|---|---|---|---|---|---|---|
| B/M | 0.155 | 0.320 | 0.478 | 0.702 | 2.496 | 1.104 | 2.340*** |
| V/P | 0.890 | 1.097 | 1.343 | 1.678 | 2.370 | 1.476 | 1.480*** |
| ME | 10846 | 7784 | 6062 | 5546 | 4686 | 6984 | -6160*** |
| beta | 1.219 | 1.168 | 1.154 | 1.117 | 1.241 | 1.186 | 0.022 |
| Ret12 | 0.136 | 0.141 | 0.153 | 0.152 | 0.177 | 0.152 | 0.041** |
| Ret24 | 0.300 | 0.314 | 0.340 | 0.334 | 0.411 | 0.340 | 0.111* |
| Ret36 | 0.386 | 0.405 | 0.450 | 0.453 | 0.542 | 0.447 | 0.156** |
| SRet12 | 0.024 | 0.027 | 0.037 | 0.033 | 0.054 | 0.035 | 0.030** |
| SRet24 | 0.042 | 0.048 | 0.069 | 0.058 | 0.118 | 0.067 | 0.076* |
| SRet36 | 0.058 | 0.066 | 0.083 | 0.101 | 0.165 | 0.094 | 0.106* |
| Obs. | 3325 | 3314 | 3315 | 3296 | 3330 | 16,580 | |

Panel C- Value-to-Price (V/P) portfolios

|  | Q1<br>Low V/P | Q2 | Q3 | Q4 | Q5<br>High V/P | All Firms | Q5-Q1<br>Diff. |
|---|---|---|---|---|---|---|---|
| V/P | 0.580 | 0.876 | 1.142 | 1.531 | 3.184 | 1.476 | 2.604*** |
| B/M | 0.538 | 0.647 | 0.858 | 1.368 | 1.800 | 1.104 | 1.261*** |
| ME | 12400 | 7747 | 5033 | 3086 | 944 | 6984 | -11455** |
| beta | 1.10 | 1.04 | 1.03 | 1.13 | 1.28 | 1.186 | 0.18 |
| Ret12 | 0.148 | 0.149 | 0.137 | 0.176 | 0.201 | 0.152 | 0.053* |
| Ret24 | 0.330 | 0.307 | 0.325 | 0.365 | 0.467 | 0.340 | 0.137** |
| Ret36 | 0.413 | 0.382 | 0.383 | 0.449 | 0.691 | 0.447 | 0.278** |
| SRet12 | 0.028 | 0.024 | 0.014 | 0.048 | 0.060 | 0.035 | 0.032* |
| SRet24 | 0.050 | 0.035 | 0.042 | 0.070 | 0.138 | 0.067 | 0.088* |
| SRet36 | 0.077 | 0.043 | 0.046 | 0.083 | 0.223 | 0.094 | 0.145* |
| Obs. | 3325 | 3314 | 3315 | 3296 | 3330 | 16,580 | |

All the NYSE, AMEX, NASDAQE stocks in the sample, were sorted into five quintile portfolios based on $ME$, $B/M$, or $V/P$ at the end of June each year. $ME$ is the market value of equity at the end of June of year $t$. $V/P$ is the fundamental value of year $t-1$, measured using the previous five years' data, divided by the stock price at the end of June of year $t$. $B/M$ is the book value of equity at the end of December of year $t-1$ divided by the market value of equity at the end of June of year $t$. The stocks in Q1 (Q5) were the stocks with the lowest (highest) $ME$, $B/M$, or $V/P$. Each panel of the table reports the characteristics of the quintile portfolios. Beta is estimated using the next 36 months of returns data. Ret12, Ret24 and Ret36 are the average buy-and-hold returns over 12, 24 and 36 months, respectively, at the beginning of July of year t. Sret12, Sret24, and Sret36 are size-adjusted returns over 12, 24 and 36 months, respectively, beginning from the July of year t. The size-adjusted returns were calculated as the difference between the raw returns and the corresponding size index returns where the cut-off point of the size deciles was based on all AMEX and NYSE stocks. Obs. denotes the number of observations in each quintile portfolio and it applies to all variables except returns. Q5-Q1 diff. represents the differences between the top portfolio and bottom portfolio. The statistical significance of the difference is based on t-statistics derived from the annual mean and the standard error of the variables. The procedures of Newey and West (1987) were followed to adjust for the serial correlation for Ret24, Ret36, Sret24, Sret36 due to overlapping holding periods. *, ** and *** denote statistical significance at 10%, 5% and 1%, respectively (two-sided tests).

Table 2 Pearson and Spearman correlation matrix among V/P and various risk indicators

|          | V/P       | Analysts  | Ln(ME)    | D/M       | Beta      | volatility | Std. ROA  | Z Score   | B/M       |
|----------|-----------|-----------|-----------|-----------|-----------|------------|-----------|-----------|-----------|
| V/P      | -         | -0.275*** | -0.478*** | 0.161***  | 0.118***  | 0.437***   | 0.112***  | -0.017    | 0.148***  |
| Analysts | -0.391*** | -         | 0.721***  | -0.125*** | -0.053*** | -0.270***  | -0.075*** | -0.157*** | -0.159*** |
| Ln(ME)   | -0.553*** | 0.754***  | -         | -0.139*** | -0.103*** | -0.491***  | -0.115*** | -0.173*** | -0.165*** |
| D/M      | 0.196***  | 0.039***  | 0.086***  | -         | 0.034***  | 0.035***   | -0.038*** | -0.146*** | 0.791***  |
| Beta     | 0.116***  | -0.032*** | -0.092*** | -0.059*** | -         | 0.2912***  | 0.136***  | 0.015*    | 0.021***  |
| Volatility | 0.361*** | -0.327*** | -0.538*** | -0.178*** | 0.290***  | -          | 0.148***  | 0.101***  | -0.006    |
| Std. ROA | 0.154***  | -0.135*** | -0.222*** | -0.223*** | 0.286***  | 0.369***   | -         | 0.011***  | -0.035*** |
| Z Score  | -0.039    | -0.176*** | -0.201*** | -0.341*** | 0.027***  | 0.130***   | 0.079***  | -         | -0.103*** |
| B/M      | 0.577***  | -0.237*** | -0.303*** | 0.431***  | 0.025***  | 0.037***   | -0.069*** | -0.218*** | -         |

The table reports the Pearson (Spearman) correlation matrix over (under) the diagonal. $V/P$ is the fundamental value of year $t-1$, measured using the previous five years' data, divided by the stock price at the end of June in year $t$. $Analysts$ is the number of financial analysts following the share, which is included in I/B/E/S files in the month following the annual earnings announcements. Ln(ME) is the logarithm of the market value of equity at the end of June in year $t$. D/M is the ratio of the long-term debt at the end of December of year t-1 to the market value of equity as measured at the end of June in year $t$. Beta is measured using the capital asset pricing model (CAPM) using monthly data over the maximum of 36 previous months ending in the December of year $t$-1. Volatility is calculated as the standard deviation of the residual from the CAPM model using daily returns data that end on the last trading day of year t-1. Std. ROA is the standard deviation of return over assets over the previous five years ending in the December of year t-1. Z score is an Altman's Z score (1968) calculated as $ALtman's\ Z$ = 0/012*(working capital/Total assets) + 0.014*(retained earnings/Total assets) + 0.033*(earnings before interest and tax/Total assets) + 0.006*(Market value of equity/Book value of Total liabilities) + 0.999*(Sales/Total assets). B/M is calculated as the book value of equity of year t-1 divided by the market value of equity at the end of June in year t. *, ** and *** denote statistical significance at 10%, 5% and 1%, respectively (two-sided tests).

Table 3 Regression of V/P on various risk factors

$$V/P = \beta_0 + \beta_1 Beta + \beta_2 Ivolatility + \beta_3\, D/M + \beta_4 Ln(ME) + \beta_5 Analysts + \beta_6 Altman's\ Z + \beta_7 Std(ROA) + \beta_8\, B/M + \varepsilon$$

Equation 3

|  | Model I | | Model II | |
| --- | --- | --- | --- | --- |
|  | $\beta$ | t-statistics | $\beta$ | t-statistics |
| Intercept | 3.312 | 40.60 *** | 2.55 | 26.58*** |
| Beta | 0.144 | 8.72*** | -0.046 | -6.92*** |
| Ivolatility |  |  | 3.99 | 17.32*** |
| D/M |  |  | 0.041 | 2.30** |
| Ln(ME) | -0.329 | -49.03*** | -0.28 | -29.55*** |
| Analysts |  |  | 0.018 | 11.23*** |
| Altman's Z |  |  | -0.167 | -9.95*** |
| Std(ROA) |  |  | 1.33 | 6.92*** |
| B/M | 0.027 | 6.51*** | 0.018 | 2.18** |
| Industry dummy | Yes |  | Yes |  |
| Year dummy | Yes |  | Yes |  |
| Adj. $R^2$ | 26.5% |  | 33.17% |  |
| Obs. | 16548 |  | 16548 |  |
| Years | 1993-2014 |  | 1992-2014 |  |

The table reports the pooled regression of Equation 3 with year and industry fixed effect. The industry classification is based on Fama and French (1997), as reported in Table 1. $V/P$ is the fundamental value of year $t-1$ divided by the stock price at the end of June in year $t$. $Analysts$ is the number of financial analysts following the share. Ln(ME) is the logarithm of the market value of equity at the end of June in year $t$. D/M is the ratio of the long-term debt at the end of December in year t-1 to the market value of equity as measured at the end of June in year $t$. Beta is measured using the capital asset pricing model (CAPM) using monthly data over the maximum of 36 previous months ending at the December of year $t$-1. Volatility is calculated as the standard deviation of the residual from the CAPM model using daily returns data ending on the last trading day of year-1. Std( ROA) is the standard deviation of return over assets over the previous five years ending in the December of year t-1. Z score is an Altman's Z score (1968). B/M is calculated as the book value of equity of year t-1 divided by the market value of equity at the end of June of year t. Adj. $R^2$ is the adjusted R-square. *, ** and *** denote statistical significance at 10%, 5% and 1%, respectively (two-sided tests)

47

Table 4 Regression of future returns on V/P and various risk factors

$$\text{Ret36} = \beta_0 + \beta_1 V/P + \beta_2 Beta + \beta_3 Ivolatility + \beta_4 D/M + \beta_5 Ln(ME) + \beta_6 Analysts + \beta_7 Altman's Z + \beta_8 Std(ROA) + \beta_9 B/M + \varepsilon$$

Equation 4

|  | V/P only | | Model I | | Model II | |
| --- | --- | --- | --- | --- | --- | --- |
|  | β | t-statistics | β | t-statistics | β | t-statistics |
| Intercept | 0.522 | 9.24*** | 0.852 | 12.62*** | 0.647 | 7.94*** |
| V/P | 0.054 | 10.23*** | 0.031 | 4.67*** | 0.013 | 1.9** |
| Beta |  |  | 0.009 | 0.93 | -0.030 | -2.58*** |
| Ivolatility |  |  |  |  | 1.431 | 7.49*** |
| D/M |  |  |  |  | 0.036 | 4.79*** |
| Ln(ME) |  |  | -0.428 | -9.35*** | -0.037 | -5.06*** |
| Analysts |  |  |  |  | 0.003 | 2.51** |
| Altman's Z |  |  |  |  | 0.045 | 3.54*** |
| Std(ROA) |  |  |  |  | -0.163 | -1.29 |
| B/M |  |  | -0.003 | -1.23 | -0.014 | -3.25*** |
| Ind. dummy | Yes |  | Yes |  | Yes |  |
| Year dummy | Yes |  | Yes |  | Yes |  |
| Adj. $R^2$ | 21.88% |  | 22.50% |  | 23.39% |  |
| Obs. | 12733 |  | 12707 |  | 11693 |  |

The table reports the pooled regression of Equation 3 with year and industry fixed effect. The industry classification is based on Fama and French (1997), as reported in Table 1. $V/P$ is the fundamental value of year $t-1$ divided by the stock price at the end of June in year $t$. $Analysts$ is the number of financial analysts following the share. Ln(ME) is the logarithm of the market value of equity at the end of June in year $t$. D/M is the ratio of the long-term debt at the end of December in year t-1 to the market value of equity as measured at the end of June in year $t$. Beta is measured using the capital asset pricing model (CAPM) using monthly data over the maximum of 36 previous months ending at the December of year $t$-1. Volatility is calculated as the standard deviation of the residual from the CAPM model using daily returns data ending on the last trading day of year-1. Std( ROA) is the standard deviation of return over assets over the previous five years ending in the December of year t-1. Z score is an Altman's Z score (1968). B/M is calculated as the book value of equity of year t-1 divided by the market value of equity at the end of June of year t. Adj. R2 is the adjusted R-square. *, ** and *** denote statistical significance at 10%, 5% and 1%, respectively (two-sided tests)

Table 5. One-factor, three-factor and five-factor models' regression for quintile portfolios formed on (V/P)

**Panel A: One factor model**

$$(R_{it} - R_{ft}) = \alpha_i + \beta_i(R_{mt} - R_{ft}) + \varepsilon_t \quad (1)$$

| | First year returns | | | | | Overlapping returns | | | | |
|---|---|---|---|---|---|---|---|---|---|---|
| | Q1 | Q2 | Q3 | Q4 | Q5 | Q1 | Q2 | Q3 | Q4 | Q5 |
| $\alpha$ | 0.001 | 0.001 | 0.001 | 0.001 | 0.004* | 0.002* | 0.002 | 0.001 | 0.002 | 0.006*** |
| | (1.14) | (1.16) | (0.66) | (1.01) | (1.88) | (1.89) | (1.62) | (1.33) | (1.51) | (2.81) |
| $\beta$ | 1.094*** | 1.095*** | 1.007*** | 1.126*** | 1.269*** | 1.093*** | 1.050*** | 0.968*** | 1.083*** | 1.174*** |
| | (39.37) | (34.49) | (26.60) | (25.91) | (21.31) | (42.81) | (35.00) | (28.97) | (27.13) | (22.93) |
| Adj.$R^2$ | 0.851 | 0.812 | 0.719 | 0.709 | 0.622 | 0.869 | 0.816 | 0.753 | 0.727 | 0.656 |
| | **GRS F-value: 4.55** | | | | | **GRS F-value: 11.450*** | | | | |

**Panel B: Three-factor model**

$$(R_{it} - R_{ft}) = \alpha_i + \beta_i(R_{mt} - R_{ft}) + s_i SMB_t + h_i HML_t + \varepsilon_t \quad (2)$$

| | First year returns | | | | | Overlapping returns | | | | |
|---|---|---|---|---|---|---|---|---|---|---|
| | Q1 | Q2 | Q3 | Q4 | Q5 | Q1 | Q2 | Q3 | Q4 | Q5 |
| $\alpha$ | 0.001 | 0.000 | 0.000 | 0.000 | 0.003 | 0.001* | 0.001 | 0.000 | 0.000 | 0.004*** |
| | (1.24) | (0.54) | (0.32) | (0.15) | (1.53) | (1.81) | (0.98) | (0.45) | (0.75) | (2.78) |
| $\beta$ | 1.050*** | 1.088*** | 1.015*** | 1.097*** | 1.200*** | 1.057*** | 1.048*** | 0.969*** | 1.063*** | 1.116*** |
| | (40.39) | (38.32) | (33.03) | (33.85) | (25.35) | (44.85) | (40.85) | (38.68) | (38.36) | (29.15) |
| $s$ | 0.244*** | 0.277*** | 0.327*** | 0.578*** | 0.831*** | 0.270*** | 0.272*** | 0.355*** | 0.537*** | 0.757*** |
| | (6.76) | (7.03) | (7.68) | (12.86) | (12.65) | (8.27) | (7.64) | (10.22) | (13.96) | (14.25) |
| $h$ | -0.078** | 0.292*** | 0.494*** | 0.489*** | 0.446*** | 0.027 | 0.329*** | 0.462*** | 0.518*** | 0.449*** |
| | (2.12) | (7.22) | (11.29) | (10.60) | (6.62) | (0.81) | (9.00) | (12.95) | (13.13) | (8.24) |
| Adj.$R^2$ | 0.875 | 0.858 | 0.825 | 0.847 | 0.774 | 0.894 | 0.873 | 0.869 | 0.876 | 0.818 |
| | **GRS F-value: 5.49** | | | | | **GRS F-value: 13.23*** | | | | |

**Panel C: Five-factor model**

$$(R_{it} - R_{ft}) = \alpha_i + \beta_i(R_{mt} - R_{ft}) + s_i SMB_t + h_i HML_t + r_i RMW_t + c_i CMA_t + \varepsilon_t \quad (3)$$

| | First year returns | | | | | Overlapping returns | | | | |
|---|---|---|---|---|---|---|---|---|---|---|
| | Q1 | Q2 | Q3 | Q4 | Q5 | Q1 | Q2 | Q3 | Q4 | Q5 |
| $\alpha$ | 0.001 | 0.000 | -0.002** | -0.001 | 0.003 | 0.001 | 0.000 | -0.001 | 0.000 | 0.004** |
| | (0.96) | (0.64) | (-2.03) | (-1.06) | (1.41) | (1.54) | (0.48) | (-1.63) | (0.49) | (2.34) |
| $\beta$ | 1.062*** | 1.159*** | 1.123*** | 1.181*** | 1.205*** | 1.067*** | 1.126*** | 1.074*** | 1.135*** | 1.141*** |
| | (33.88) | (34.70) | (32.06) | (31.03) | (21.12) | (37.56) | (37.98) | (38.76) | (34.92) | (24.77) |
| $s$ | 0.249*** | 0.331*** | 0.425*** | 0.633*** | 0.810*** | 0.267*** | 0.348*** | 0.458*** | 0.600*** | 0.761*** |
| | (6.13) | (7.62) | (9.33) | (12.80) | (10.92) | (7.23) | (9.02) | (12.71) | (14.21) | (12.69) |
| $h$ | -0.106** | 0.152*** | 0.295*** | 0.315*** | 0.410*** | -0.002 | 0.191*** | 0.277*** | 0.386*** | 0.381*** |
| | (-2.01) | (2.70) | (4.97) | (4.90) | (4.25) | (-0.06) | (3.82) | (5.91) | (7.02) | (4.89) |
| $r$ | 0.0261 | 0.196*** | 0.336*** | 0.212*** | -0.043 | 0.002 | 0.254*** | 0.344*** | 0.219*** | 0.035 |
| | (0.46) | (3.25) | (5.30) | (3.09) | (-0.42) | (0.04) | (4.75) | (6.87) | (3.73) | (0.43) |
| $c$ | 0.046 | 0.185*** | 0.223*** | 0.252*** | 0.113 | 0.068 | 0.137** | 0.184** | 0.151** | 0.131 |
| | (0.63) | (2.37) | (2.72) | (2.83) | (0.85) | (1.02) | (1.98) | (2.84) | (1.99) | (1.22) |
| Adj.$R^2$ | 0.874 | 0.864 | 0.843 | 0.854 | 0.773 | 0.894 | 0.883 | 0.889 | 0.882 | 0.818 |
| | **GRS F-value: 16.88**** | | | | | **GRS F-value: 29.51**** | | | | |

The table reports the regression results of the CAPM and Fama and French's three- and five-factor models by regressing the excess monthly returns of the V/P quintile portfolios against the market excess returns and combinations of SMB, HML, CMA and RMW factors. The performance of different models was compared by calculating the GRS F-statistics. The significant GRS F-statistics indicate that the combined intercepts are not equal to zero. All the NYSE, AMEX, NASDAQE stocks in the sample, were sorted into five quintile portfolios based on V/P at the end of June each year. V/P is the fundamental value at the end of December of year $t-1$, measured using the previous five years' data, divided by the stock price at the end of June of year $t$. Q1 consists of stocks with the lowest V/P ratio and Q5 consists of stocks with the highest V/P ratio. $R_{it}$ are the monthly returns of the quintile portfolio $i$. $R_{mt}$ are the monthly returns of the market index. $R_{ft}$ denotes the monthly riskless rate on treasury bills. SMB is the size factor and is measured as the returns on small stocks minus the returns on the large stocks. HML is the book to market ratio factor and is measured as the returns on shares with a high B/M ratio minus the return on stocks with a low B/M ratio. RMW is the profitability factor and is measured as the return on the robust stocks (top 30% of shares with the highest operating income) minus the return on the weak stocks (bottom 30% of shares with the smallest operating income). CMA is the investment factor whose returns make up the difference between the conservative stocks (top 30% of firms with the smallest changes in total assets) and the returns on the most aggressive stocks (lowest 30% of firms with the greatest change in total assets). The monthly data for $R_{mt}$, $R_{ft}$, SMB, HML, RMW and CMA were downloaded from Ken French's website of data for 276 months over the period from 1993 to 2015. *, ** and *** denote significant level at 10%, 5% and 1%, respectively (for two-sided tests). Overlapping returns, in any given month t, is calculated by holding a series of portfolios that are selected in the current year as well as in the previous year (t-1) and the year before that (t-2). Specifically, overlapping returns for each quantile portfolio (Q1- Q5) consists of 1/3 of the first year returns for each quantile portfolio formed in year t, and 1/3 of the second year returns for each quantile portfolio formed in year t-1 and 1/3 of the third year returns for each quantile portfolios formed in year t-2.

Table 6. The effect of second and third year returns for quantile portfolios formed on (V/P)

**Panel A: One-factor model**

$$(R_{it} - R_{ft}) = \alpha_i + \beta_i(R_{mt} - R_{ft}) + \varepsilon_t \quad (1)$$

|  | Second year returns | | | | | Third year returns | | | | |
|---|---|---|---|---|---|---|---|---|---|---|
|  | Q1 | Q2 | Q3 | Q4 | Q5 | Q1 | Q2 | Q3 | Q4 | Q5 |
| $\alpha$ | 0.001 | 0.001 | 0.002* | 0.003* | 0.006*** | 0.003** | 0.002 | 0.002 | 0.003* | 0.007*** |
|  | (1.51) | (1.39) | (1.78) | (1.71) | (2.76) | (2.21) | (1.45) | (1.47) | (1.72) | (3.31) |
| $\beta$ | 1.109*** | 1.033*** | 0.962*** | 1.090*** | 1.138*** | 1.078*** | 1.022*** | 0.931*** | 1.046*** | 1.128*** |
|  | (37.55) | (32.07) | (26.50) | (25.80) | (20.58) | (33.35) | (29.41) | (26.97) | (23.90) | (21.30) |
| Adj.$R^2$ | **0.842** | **0.796** | **0.727** | **0.716** | **0.616** | **0.815** | **0.774** | **0.743** | **0.694** | **0.643** |
|  | GRS F-value: 8.844 | | | | | GRS F-value: 14.78** | | | | |

**Panel B: Three-factor model**

$$(R_{it} - R_{ft}) = \alpha_i + \beta_i(R_{mt} - R_{ft}) + s_i SMB_t + h_i HML_t + \varepsilon_t \quad (2)$$

|  | Second year returns | | | | | Third year returns | | | | |
|---|---|---|---|---|---|---|---|---|---|---|
|  | Q1 | Q2 | Q3 | Q4 | Q5 | Q1 | Q2 | Q3 | Q4 | Q5 |
| $\alpha$ | 0.001 | 0.000 | 0.001 | 0.001 | 0.005*** | 0.002** | 0.001 | 0.000 | 0.001 | 0.006*** |
|  | (1.38) | (0.81) | (1.25) | (1.20) | (2.80) | (1.99) | (0.81) | (0.70) | (1.04) | (3.27) |
| $\beta$ | 1.076*** | 1.034*** | 0.956*** | 1.070*** | 1.079*** | 1.045*** | 1.022*** | 0.930*** | 1.036*** | 1.078*** |
|  | (38.23) | (36.76) | (34.56) | (35.01) | (25.78) | (34.67) | (33.75) | (35.41) | (32.46) | (25.20) |
| $s$ | 0.262*** | 0.257*** | 0.394*** | 0.537*** | 0.783*** | 0.307*** | 0.274*** | 0.345*** | 0.491*** | 0.661*** |
|  | (6.73) | (6.60) | (10.28) | (12.69) | (13.51) | (7.37) | (6.53) | (9.49) | (11.12) | (11.15) |
| $h$ | 0.046 | 0.339*** | 0.459*** | 0.513*** | 0.476*** | 0.114*** | 0.360*** | 0.446*** | 0.563*** | 0.430*** |
|  | (1.15) | (8.50) | (11.70) | (11.84) | (8.02) | (2.67) | (8.40) | (11.99) | (12.46) | (7.10) |
| Adj.$R^2$ | **0.865** | **0.853** | **0.850** | **0.860** | **0.792** | **0.849** | **0.839** | **0.860** | **0.847** | **0.781** |
|  | GRS F-value: 9.352* | | | | | GRS F-value: 16.76** | | | | |

**Panel C: Five factor model**

$$(R_{it} - R_{ft}) = \alpha_i + \beta_i(R_{mt} - R_{ft}) + s_i SMB_t + h_i HML_t + r_i RMW_t + c_i CMA_t + \varepsilon_t \quad (3)$$

|  | Second year returns | | | | | Third year returns | | | | |
|---|---|---|---|---|---|---|---|---|---|---|
|  | Q1 | Q2 | Q3 | Q4 | Q5 | Q1 | Q2 | Q3 | Q4 | Q5 |
| $\alpha$ | 0.001 | 0.000 | 0.000 | 0.000 | 0.004*** | 0.002* | 0.000 | -0.001 | 0.000 | 0.005** |
|  | (1.42) | (0.48) | (0.56) | (0.17) | (2.37) | (1.61) | (0.59) | (1.26) | (0.02) | (2.66) |
| $\beta$ | 1.069*** | 1.108*** | 1.059*** | 1.134*** | 1.108*** | 1.067*** | 1.110*** | 1.034*** | 1.103*** | 1.121*** |
|  | (31.64) | (34.04) | (33.87) | (31.37) | (22.13) | (30.16) | (31.76) | (35.83) | (29.43) | (21.87) |
| $s$ | 0.285*** | 0.347*** | 0.488*** | 0.594*** | 0.762*** | 0.268*** | 0.359*** | 0.461*** | 0.569*** | 0.713*** |
|  | (6.48) | (8.18) | (11.97) | (12.59) | (11.66) | (5.79) | (7.84) | (12.20) | (11.60) | (10.63) |
| $h$ | 0.089 | 0.222*** | 0.268*** | 0.391*** | 0.373*** | 0.015 | 0.197*** | 0.270*** | 0.452*** | 0.363*** |
|  | (1.57) | (4.03) | (5.09) | (6.40) | (4.42) | (0.26) | (3.34) | (5.53) | (7.12) | (4.18) |
| $r$ | 0.042 | 0.281*** | 0.319*** | 0.195*** | -0.016 | -0.064 | 0.281*** | 0.365*** | 0.243*** | 0.161* |
|  | (0.70) | (4.75) | (5.61) | (2.97) | (-0.18) | (-1.01) | (4.42) | (6.96) | (3.57) | (1.73) |
| $c$ | -0.132* | 0.067 | 0.210*** | 0.142* | 0.251** | 0.277*** | 0.168** | 0.138** | 0.077 | 0.037 |
|  | (1.68) | (0.88) | (2.88) | (1.69) | (2.15) | (3.37) | (2.07) | (2.05) | (0.89) | (0.31) |
| Adj.$R^2$ | **0.865** | **0.864** | **0.868** | **0.864** | **0.795** | **0.856** | **0.851** | **0.883** | **0.853** | **0.782** |
|  | GRS F-value: 14.24** | | | | | GRS F-value: 25.51** | | | | |

The table reports the regression results of the CAPM and Fama and French's three- and five-factor models by regressing the excess monthly returns of the V/P quintile portfolios against the market excess returns and combinations of SMB, HML, CMA and RMW factors. The performance of different models was compared by calculating the GRS F-statistics. The significant GRS F-statistics indicate that the combined intercepts are not equal to zero. All the NYSE, AMEX, NASDAQE stocks in the sample, were sorted into five quintile portfolios based on V/P at the end of June each year. V/P is the fundamental value at the end of December of year t-1, measured using the previous five years' data, divided by the stock price at the end of June of year $t$. Q1 consists of stocks with the lowest V/P ratio and Q5 consists of stocks with the highest V/P ratio. $R_{it}$ are the monthly returns of the quintile portfolio $i$. $R_{mt}$ are the monthly returns of the market index. $R_{ft}$ denotes the monthly riskless rate on treasury bills. SMB is the size factor and is measured as the returns on small stocks minus the returns on the large stocks. HML is the book to market ratio factor and is measured as the returns on shares with a high B/M ratio minus the return on stocks with a low B/M ratio. RMW is the profitability factor and is measured as the return on the robust stocks (top 30% of shares with the highest operating income) minus the return on the weak stocks (bottom 30% of shares with the smallest operating income). CMA is the investment factor whose returns make up the difference between the conservative stocks (top 30% of firms with the smallest changes in total assets) and the returns on the most aggressive stocks (lowest 30% of firms with the greatest change in total assets). The monthly data for $R_{mt}$, $R_{ft}$, SMB, HML, RMW and CMA were downloaded from Ken French's website of data for 276 months over the period from 1993 to 2015. *, ** and *** denote significant level at 10%, 5% and 1%, respectively (for two-sided tests). Overlapping returns, in any given month t, is calculated by holding a series of portfolios that are selected in the current year as well as in the previous year (t-1) and the year before that (t-2). Specifically, overlapping returns for each quantile portfolio (Q1- Q5) consists of 1/3 of the first year returns for each quantile portfolio formed in year t, and 1/3 of the second year returns for each quantile portfolio formed in year t-1 and 1/3 of the third year returns for each quantile portfolios formed in year t-2.

Table 7. The effect of dividends for quintile portfolios formed on (V/P)

Panel A: One-factor model

$$(R_{it} - R_{ft}) = \alpha_i + \beta_i(R_{mt} - R_{ft}) + \varepsilon_t \quad (1)$$

| | First year returns | | | | | Overlapping returns | | | | |
|---|---|---|---|---|---|---|---|---|---|---|
| | Q1 | Q2 | Q3 | Q4 | Q5 | Q1 | Q2 | Q3 | Q4 | Q5 |
| $\alpha$ | 0.002** | 0.003** | 0.002* | 0.003* | 0.006** | 0.003*** | 0.003*** | 0.003** | 0.004** | 0.007*** |
| | (1.97) | (2.18) | (1.65) | (1.93) | (2.32) | (2.83) | (2.70) | (2.46) | (2.52) | (3.30) |
| $\beta$ | 1.094*** | 1.096*** | 1.007*** | 1.128*** | 1.269*** | 1.093*** | 1.051*** | 0.969*** | 1.085*** | 1.174** |
| | (39.68) | (34.45) | (26.54) | (25.95) | (21.33) | (42.86) | (34.93) | (28.94) | (27.22) | (22.93) |
| Adj.R2 | 0.851 | 0.811 | 0.718 | 0.709 | 0.623 | 0.869 | 0.816 | 0.752 | 0.729 | 0.656 |
| | GRS F-value: 6.703 | | | | | GRS F-value: 13.54** | | | | |

Panel B: Three-factor model

$$(R_{it} - R_{ft}) = \alpha_i + \beta_i(R_{mt} - R_{ft}) + s_i SMB_t + h_i HML_t + \varepsilon_t \quad (2)$$

| | First year returns | | | | | Overlapping returns | | | | |
|---|---|---|---|---|---|---|---|---|---|---|
| | Q1 | Q2 | Q3 | Q4 | Q5 | Q1 | Q2 | Q3 | Q4 | Q5 |
| $\alpha$ | 0.002** | 0.002* | 0.001 | 0.001 | 0.004** | 0.002*** | 0.002** | 0.002** | 0.002** | 0.005*** |
| | (2.14) | (1.71) | (0.93) | (1.42) | (2.10) | (2.85) | (2.27) | (2.00) | (2.24) | (3.46) |
| $\beta$ | 1.051*** | 1.088*** | 1.015*** | 1.099*** | 1.200*** | 1.058*** | 1.048*** | 0.969*** | 1.065*** | 1.116*** |
| | (40.42) | (38.27) | (32.96) | (33.96) | (25.42) | (44.94) | (40.78) | (38.71) | (38.60) | (29.21) |
| $s$ | 0.244*** | 0.278*** | 0.329*** | 0.580*** | 0.833*** | 0.271*** | 0.273*** | 0.357*** | 0.538*** | 0.759*** |
| | (6.78) | (7.06) | (7.71) | (12.90) | (12.71) | (8.30) | (7.67) | (10.29) | (14.04) | (14.31) |
| $h$ | -0.079** | 0.291*** | 0.495*** | 0.489*** | 0.447*** | 0.026 | 0.329*** | 0.463*** | 0.518*** | 0.450*** |
| | (-2.16) | (7.19) | (11.29) | (10.62) | (6.66) | (0.79) | (9.00) | (12.99) | (13.19) | (8.28) |
| Adj.R2 | 0.875 | 0.858 | 0.825 | 0.848 | 0.775 | 0.895 | 0.873 | 0.869 | 0.877 | 0.819 |
| | GRS F-value: 6.84 | | | | | GRS F-value: 15.04* | | | | |

Panel C: Five-factor model

$$(R_{it} - R_{ft}) = \alpha_i + \beta_i(R_{mt} - R_{ft}) + s_i SMB_t + h_i HML_t + r_i RMW_t + c_i CMA_t + \varepsilon_t \quad (3)$$

| | First year returns | | | | | Overlapping returns | | | | |
|---|---|---|---|---|---|---|---|---|---|---|
| | Q1 | Q2 | Q3 | Q4 | Q5 | Q1 | Q2 | Q3 | Q4 | Q5 |
| $\alpha$ | 0.002* | 0.000 | -0.001 | 0.000 | 0.004* | 0.002** | 0.000 | -0.000 | 0.001 | 0.005*** |
| | (1.82) | (0.50) | (-0.78) | (0.19) | (1.95) | (2.53) | (0.81) | (-0.03) | (0.98) | (2.99) |
| $\beta$ | 1.063*** | 1.159*** | 1.124*** | 1.182*** | 1.204*** | 1.067*** | 1.126*** | 1.075*** | 1.136*** | 1.141*** |
| | (33.90) | (34.64) | (32.01) | (31.10) | (21.18) | (37.64) | (37.88) | (38.80) | (35.13) | (24.82) |
| $s$ | 0.249*** | 0.331*** | 0.426*** | 0.633*** | 0.811*** | 0.267*** | 0.348*** | 0.460*** | 0.600*** | 0.761*** |
| | (6.12) | (7.63) | (9.33) | (12.81) | (10.96) | (7.25) | (9.02) | (12.76) | (14.27) | (12.73) |
| $h$ | -0.107** | 0.151*** | 0.293*** | 0.316*** | 0.411*** | -0.003 | 0.191*** | 0.276*** | 0.386*** | 0.382** |
| | (-2.03) | (2.68) | (4.94) | (4.92) | (4.28) | (-0.08) | (3.82) | (5.91) | (7.06) | (4.92) |
| $r$ | 0.024 | 0.195*** | 0.335*** | 0.208*** | -0.044 | 0.002 | 0.252*** | 0.343*** | 0.217*** | 0.033 |
| | (0.43) | (3.23) | (5.27) | (3.03) | (-0.43) | (0.04) | (4.69) | (6.85) | (3.72) | (0.40) |
| $c$ | 0.047 | 0.185** | 0.229*** | 0.254*** | 0.114 | 0.068 | 0.140** | 0.189*** | 0.152** | 0.133 |
| | (0.65) | (2.37) | (2.79) | (2.85) | (0.86) | (1.03) | (2.02) | (2.92) | (2.01) | (1.24) |
| Adj.R2 | 0.874 | 0.864 | 0.843 | 0.855 | 0.775 | 0.895 | 0.882 | 0.889 | 0.883 | 0.819 |
| | GRS F-value: 12.67* | | | | | GRS F-value: 22.84** | | | | |

The table reports the regression results of the CAPM and Fama and French's three- and five-factor models by regressing the excess monthly returns of the V/P quintile portfolios against the market excess returns and combinations of SMB, HML, CMA and RMW factors. The performance of different models was compared by calculating the GRS F-statistics. The significant GRS F-statistics indicate that the combined intercepts are not equal to zero. All the NYSE, AMEX, NASDAQE stocks in the sample, were sorted into five quintile portfolios based on V/P at the end of June each year. V/P is the fundamental value at the end of December of year t-1, measured using the previous five years' data, divided by the stock price at the end of June of year $t$. Q1 consists of stocks with the lowest V/P ratio and Q5 consists of stocks with the highest V/P ratio. $R_{it}$ are the monthly returns of the quintile portfolio $i$. $R_{mt}$ are the monthly returns of the market index. $R_{ft}$ denotes the monthly riskless rate on treasury bills. SMB is the size factor and is measured as the returns on small stocks minus the returns on the large stocks. HML is the book to market ratio factor and is measured as the returns on shares with a high B/M ratio minus the return on stocks with a low B/M ratio. RMW is the profitability factor and is measured as the return on the robust stocks (top 30% of shares with the highest operating income) minus the return on the weak stocks (bottom 30% of shares with the smallest operating income). CMA is the investment factor whose returns make up the difference between the conservative stocks (top 30% of firms with the smallest changes in total assets) and the returns on the most aggressive stocks (lowest 30% of firms with the greatest change in total assets). The monthly data for $R_{mt}$, $R_{ft}$, SMB, HML, RMW and CMA were downloaded from Ken French's website of data for 276 months over the period from 1993 to 2015. *, ** and *** denote significant level at 10%, 5% and 1%, respectively (for two-sided tests). Overlapping returns, in any given month t, is calculated by holding a series of portfolios that are selected in the current year as well as in the previous year (t-1) and the year before that (t-2). Specifically, overlapping returns for each quantile portfolio (Q1- Q5) consists of 1/3 of the first year returns for each quantile portfolio formed in year t, and 1/3 of the second year returns for each quantile portfolio formed in year t-2 and 1/3 of the third year returns for each quantile portfolios formed in year t-3.

Table 8. The effect of dividends on second and third year returns for quintile portfolios formed on (V/P)

**Panel A: One-factor model**

$$(R_{it} - R_{ft}) = \alpha_i + \beta_i(R_{mt} - R_{ft}) + \varepsilon_t \quad (1)$$

|  | Second year returns | | | | | Third year returns | | | | |
|---|---|---|---|---|---|---|---|---|---|---|
|  | Q1 | Q2 | Q3 | Q4 | Q5 | Q1 | Q2 | Q3 | Q4 | Q5 |
| $\alpha$ | 0.003** | 0.003** | 0.004*** | 0.004*** | 0.007*** | 0.004*** | 0.003** | 0.004** | 0.005*** | 0.008*** |
|  | (2.27) | (2.36) | (2.82) | (2.61) | (3.20) | (2.93) | (2.33) | (2.51) | (2.58) | (3.73) |
| $\beta$ | 1.110*** | 1.034*** | 0.963*** | 1.093*** | 1.138*** | 1.079*** | 1.023*** | 0.932*** | 1.048*** | 1.129*** |
|  | (37.52) | (31.73) | (26.49) | (25.88) | (20.53) | (33.41) | (29.33) | (26.97) | (23.98) | (21.33) |
| Adj.R2 | 0.842 | 0.795 | 0.727 | 0.717 | 0.615 | 0.816 | 0.773 | 0.743 | 0.695 | 0.643 |
|  | GRS F-value: 11.97* | | | | | GRS F-value: 12.63* | | | | |

**Panel B: Three-factor model**

$$(R_{it} - R_{ft}) = \alpha_i + \beta_i(R_{mt} - R_{ft}) + s_i SMB_t + h_i HML_t + \varepsilon_t \quad (2)$$

|  | Second year returns | | | | | Third year returns | | | | |
|---|---|---|---|---|---|---|---|---|---|---|
|  | Q1 | Q2 | Q3 | Q4 | Q5 | Q1 | Q2 | Q3 | Q4 | Q5 |
| $\alpha$ | 0.002** | 0.002** | 0.003*** | 0.003** | 0.006*** | 0.003*** | 0.002* | 0.002** | 0.003** | 0.007*** |
|  | (2.20) | (1.96) | (2.65) | (2.48) | (3.40) | (2.78) | (1.85) | (2.10) | (2.26) | (3.81) |
| $\beta$ | 1.077*** | 1.034** | 0.957*** | 1.072*** | 1.078*** | 1.046*** | 1.023*** | 0.931*** | 1.038** | 1.079*** |
|  | (38.19) | (36.76) | (34.65) | (35.18) | (25.74) | (34.75) | (33.65) | (35.54) | (32.71) | (25.29) |
| $s$ | 0.263*** | 0.259*** | 0.396*** | 0.538*** | 0.786*** | 0.308*** | 0.276*** | 0.348*** | 0.493*** | 0.662*** |
|  | (6.74) | (6.65) | (10.35) | (12.74) | (13.55) | (7.40) | (6.56) | (9.61) | (11.21) | (11.20) |
| $h$ | 0.045 | 0.342*** | 0.461*** | 0.514*** | 0.478*** | 0.115*** | 0.361*** | 0.447*** | 0.563*** | 0.431*** |
|  | (1.13) | (8.55) | (11.76) | (11.89) | (8.04) | (2.69) | (8.38) | (12.06) | (12.54) | (7.14) |
| Adj.R2 | 0.864 | 0.853 | 0.851 | 0.861 | 0.792 | 0.850 | 0.838 | 0.861 | 0.849 | 0.782 |
|  | GRS F-value: 12.86* | | | | | GRS F-value: 13.83* | | | | |

**Panel F: Five-factor model**

$$(R_{it} - R_{ft}) = \alpha_i + \beta_i(R_{mt} - R_{ft}) + s_i SMB_t + h_i HML_t + r_i RMW_t + c_i CMA_t + \varepsilon_t \quad (3)$$

|  | Second year returns | | | | | Third year returns | | | | |
|---|---|---|---|---|---|---|---|---|---|---|
|  | Q1 | Q2 | Q3 | Q4 | Q5 | Q1 | Q2 | Q3 | Q4 | Q5 |
| $\alpha$ | 0.002** | 0.000 | 0.000 | 0.001 | 0.005*** | 0.003** | 0.000 | 0.000 | 0.001 | 0.006*** |
|  | (2.21) | (0.65) | (0.85) | (1.40) | (2.95) | (2.38) | (0.45) | (0.19) | (1.21) | (3.18) |
| $\beta$ | 1.069*** | 1.109*** | 1.060*** | 1.137*** | 1.094*** | 1.068*** | 1.110** | 1.034*** | 1.104*** | 1.121*** |
|  | (31.61) | (34.01) | (33.99) | (31.54) | (22.10) | (30.25) | (31.61) | (35.97) | (29.62) | (21.93) |
| $s$ | 0.286*** | 0.348*** | 0.489*** | 0.595*** | 0.763*** | 0.268*** | 0.359** | 0.463*** | 0.569*** | 0.713*** |
|  | (6.42) | (8.19) | (12.03) | (12.65) | (11.68) | (5.81) | (7.82) | (12.30) | (11.65) | (10.65) |
| $h$ | 0.088 | 0.223*** | 0.268*** | 0.391*** | 0.372*** | 0.015 | 0.198** | 0.270*** | 0.452*** | 0.364*** |
|  | (1.60) | (4.04) | (5.09) | (6.42) | (4.43) | (0.26) | (3.33) | (5.55) | (7.17) | (4.21) |
| $r$ | 0.042 | 0.279*** | 0.319*** | 0.196*** | -0.011 | -0.064 | 0.277** | 0.364*** | 0.239*** | 0.158* |
|  | (0.68) | (4.70) | (5.61) | (2.99) | (-0.23) | (-1.01) | (4.33) | (6.95) | (3.52) | (1.70) |
| $c$ | -0.131* | 0.073 | 0.215*** | 0.143* | 0.237** | 0.279*** | 0.171* | 0.141** | 0.081 | 0.038 |
|  | (-1.66) | (0.95) | (2.96) | (1.70) | (2.17) | (3.39) | (2.10) | (2.11) | (0.93) | (0.32) |
| Adj.R2 | 0.865 | 0.864 | 0.869 | 0.865 | 0.795 | 0.856 | 0.850 | 0.884 | 0.855 | 0.783 |
|  | GRS F-value: 12.94* | | | | | GRS F-value: 20.77** | | | | |

The table reports the regression results of the CAPM and Fama and French's three- and five-factor models by regressing the excess monthly returns of the V/P quintile portfolios against the market excess returns and combinations of SMB, HML, CMA and RMW factors. The performance of different models was compared by calculating the GRS F-statistics. The significant GRS F-statistics indicate that the combined intercepts are not equal to zero. All the NYSE, AMEX, NASDAQE stocks in the sample, were sorted into five quintile portfolios based on V/P at the end of June each year. V/P is the fundamental value at the end of December of year t-1, measured using the previous five years' data, divided by the stock price at the end of June of year $t$. Q1 consists of stocks with the lowest V/P ratio and Q5 consists of stocks with the highest V/P ratio. $R_{it}$ are the monthly returns of the quintile portfolio $i$. $R_{mt}$ are the monthly returns of the market index. $R_{ft}$ denotes the monthly riskless rate on treasury bills. SMB is the size factor and is measured as the returns on small stocks minus the returns on the large stocks. HML is the book to market ratio factor and is measured as the returns on shares with a high B/M ratio minus the return on stocks with a low B/M ratio. RMW is the profitability factor and is measured as the return on the robust stocks (top 30% of shares with the highest operating income) minus the return on the weak stocks (bottom 30% of shares with the smallest operating income). CMA is the investment factor whose returns make up the difference between the conservative stocks (top 30% of firms with the smallest changes in total assets) and the returns on the most aggressive stocks (lowest 30% of firms with the greatest change in total assets). The monthly data for $R_{mt}$, $R_{ft}$, SMB, HML, RMW and CMA were downloaded from Ken French's website of data for 276 months over the period from 1993 to 2015. *, ** and *** denote significant level at 10%, 5% and 1%, respectively (for two-sided tests). Overlapping returns, in any given month t, is calculated by holding a series of portfolios that are selected in the current year as well as in the previous year (t-1) and the year before that (t-2). Specifically, overlapping returns for each quantile portfolio (Q1- Q5) consists of 1/3 of the first year returns for each quantile portfolio formed in year t, and 1/3 of the second year returns for each quantile portfolio formed in year t-2 and 1/3 of the third year returns for each quantile portfolios formed in year t-3.



\end{document}